\documentclass[10pt, twocolumn]{IEEEtran}
\usepackage{color}
\usepackage{cite}
\usepackage{textcase}
\usepackage{amsmath}
\usepackage{amsthm}
\usepackage{amsfonts}
\usepackage{amssymb}
\usepackage{algorithm}
\usepackage{algorithmic}
\usepackage{graphicx}
\usepackage{verbatim}

\newtheorem{theorem}{Theorem}
\newtheorem{lemma}{Lemma}

\newtheorem{corollary}{Corollary}
\newtheorem{example}{Example}

\newcommand{\beq}{\begin{equation}}
\newcommand{\eeq}{\end{equation}}
\newcommand{\beqnn}{\begin{equation*}}
\newcommand{\eeqnn}{\end{equation*}}
\newcommand{\beqy}{\begin{eqnarray}}
\newcommand{\eeqy}{\end{eqnarray}}
\newcommand{\beqynn}{\begin{eqnarray*}}
\newcommand{\eeqynn}{\end{eqnarray*}}
\newcommand{\bit}{\begin{itemize}}
\newcommand{\eit}{\end{itemize}}
\newcommand{\ben}{\begin{enumerate}}
\newcommand{\een}{\end{enumerate}}
\newcommand{\bex}{\begin{example}}
\newcommand{\eex}{\end{example}}

\newcommand{\balg}[1]{\begin{algorithm} \caption{#1}}
\newcommand{\ealg}{\end{algorithm}}

\newcommand{\balgc}{\begin{algorithmic}[1]}
\newcommand{\ealgc}{\end{algorithmic}}

\newcommand{\bary}{\begin{array}}
\newcommand{\eary}{\end{array}}
\newcommand{\bmx}{\begin{bmatrix}}
\newcommand{\emx}{\end{bmatrix}}
\newcommand{\bsmx}{\left[\begin{smallmatrix}}
\newcommand{\esmx}{\end{smallmatrix}\right]}
\newcommand{\bmxc}[1]{\left[\begin{array}{@{}#1@{}}}
\newcommand{\emxc}{\end{array}\right]}
\newcommand{\bcn}{\begin{center}}
\newcommand{\ecn}{\end{center}}

\newcommand{\Rbb}{{\mathbb{R}}}
\newcommand{\Zbb}{{\mathbb{Z}}}

\newcommand{\Rn}{\Rbb^{n}}

\newcommand{\Rnbn}{\Rbb^{n \times n}}

\newcommand{\Rmbn}{\Rbb^{m \times n}}

\newcommand{\Zn}{\Zbb^{n}}

\newcommand{\sss}{\scriptscriptstyle }

\newcommand{\sOL}{{\scriptscriptstyle \mathrm{OL}}}
\newcommand{\sOB}{{\scriptscriptstyle \mathrm{OB}}}
\newcommand{\sOR}{{\scriptscriptstyle \mathrm{OR}}}

\newcommand{\sD}{{\scriptscriptstyle \mathrm{D}}}
\newcommand{\sR}{{\scriptscriptstyle \mathrm{R}}}
\newcommand{\sBB}{{\scriptscriptstyle \mathrm{BB}}}
\newcommand{\sBR}{{\scriptscriptstyle \mathrm{BR}}}
\newcommand{\sBL}{{\scriptscriptstyle \mathrm{BL}}}

\newcommand{\A}{\boldsymbol{A}}

\newcommand{\I}{\boldsymbol{I}}

\newcommand{\Q}{\boldsymbol{Q}}
\newcommand{\R}{\boldsymbol{R}}

\newcommand{\T}{\boldsymbol{T}}
\newcommand{\U}{\boldsymbol{U}}

\newcommand{\W}{\boldsymbol{W}}

\renewcommand{\d}{\boldsymbol{d}}
\newcommand{\e}{\boldsymbol{e}}

\renewcommand{\l}{\boldsymbol{\ell}}

\renewcommand{\u}{\boldsymbol{u}}

\renewcommand{\v}{\boldsymbol{v}}

\newcommand{\x}{{\boldsymbol{x}}}
\newcommand{\y}{{\boldsymbol{y}}}

\newcommand{\0}{{\boldsymbol{0}}}

\newcommand{\bbx}{{\bar{\x}}}

\newcommand{\ty}{{\tilde{y}}}

\newcommand{\tby}{{\tilde{\y}}}

\newcommand{\hx}{{\hat{x}}}

\newcommand{\hbx}{{\hat{\x}}}

\newcommand{\bSigma}{{\boldsymbol{\Sigma}}}
\newcommand{\bxi}{\boldsymbol{\xi}}
\newcommand{\boldeta}{\boldsymbol{\eta}}

\usepackage{slashbox}

\begin{document}

\title{On the Success Probability of Three Detectors for the Box-Constrained Integer Linear Model}

\author{Jinming~Wen and Xiao-Wen~Chang
\thanks{This work was presented in part at the 2017 IEEE International Symposium on Information Theory (ISIT), Aachen, Germany.}
\thanks{J.~Wen is with  the College of Information Science and Technology and the College of Cyber Security, Jinan University,Guangzhou, 510632, China,
J.~Wen is also with Pazhou Lab, Guangzhou, 510330, China (e-mail: jinming.wen@mail.mcgill.ca). }
\thanks{X.-W. Chang is with the School of Computer Science, McGill University,
Montreal, QC H3A 2A7, Canada (e-mail: chang@cs.mcgill.ca).}
\thanks{This work was supported by NSERC of Canada grant 217191-17, NSFC (Nos. 11871248, 61932011),
the Guangdong Province Universities and Colleges Pearl River Scholar Funded Scheme (2019),
Guangdong Major Project of  Basic and Applied Basic Research (2019B030302008, 2019B1515120010),
Natural Science Foundation of Guangdong Province of China (2021A515010857),
Guangdong Key RD Plan 2020 (No. 2020B0101090002).}
}

\maketitle

\begin{abstract}
This paper is concerned with  detecting an integer parameter vector inside a box
from a  linear model that is corrupted with a noise vector following the
Gaussian distribution. One of the commonly used detectors is the maximum
likelihood detector, which is obtained by solving a box-constrained
integer least squares problem, that is NP-hard.
Two other popular detectors are the box-constrained rounding
and Babai detectors due to their high efficiency of implementation.
In this paper, we first present formulas for the success probabilities
(the probabilities of correct detection) of these three detectors
for two different situations: the integer parameter vector is  deterministic
and is uniformly distributed over the constraint box. Then, we give two simple examples
to respectively show that the success probability of the box-constrained
rounding detector can be  larger than that of the box-constrained Babai detector and
the latter can be larger than the success probability of the maximum likelihood detector
when the parameter vector is deterministic, and prove that the success probability
of the box-constrained rounding detector is always  not larger
than that of the box-constrained Babai detector when the parameter vector is
uniformly distributed over the constraint box.
Some relations between the results for the box constrained and ordinary cases
are presented, and two bounds on the success probability of the maximum
likelihood detector, which can easily be computed, are developed.
Finally, simulation results are provided to illustrate our main theoretical findings.
\end{abstract}


\section{Introduction}
\label{s:introduction}
Suppose that we have the following box-constrained linear model:
\begin{align}
& \y=\A\hbx+\v, \quad \v \sim \mathcal{N}(\boldsymbol{0},\sigma^2 \I) \label{e:model}, \\
& \hbx\in \mathcal{B} \equiv \{\x \in \mathbb{Z}^{n}:\l \leq \x\leq \u,\; \l, \u\in \mathbb{Z}^{n}, \l<\u\},
\label{e:box}
\end{align}
where $\y\in \mathbb{R}^m$ is an observation vector,
$\A\in\mathbb{R}^{m\times n}$ with $m\geq n$ is a deterministic full column rank model matrix,
$\hat\x\in \mathbb{R}^n$ is an integer parameter vector and
$\v\in \mathbb{R}^m$ is a noise vector following the Gaussian distribution
$\mathcal{N}(\boldsymbol{0},\sigma^2 \I)$ with given $\sigma$.

This paper studies the detection of $\hbx$, which can be deterministic or random in the box $\mathcal{B}$, from \eqref{e:model}.
This problem arises from many applications.
For example, in  some wireless communications systems (see e.g., \cite{DamGC03,GuoN06, BouV98}),
$\hbx$ is a random vector which is uniformly distributed over the box $\mathcal{B}$;
in transportation science (see, e.g., \cite{KuuSS14}) and  image processing (see, e.g., \cite{BlaBS13}),
$\hbx$ is a deterministic vector in a finite box $\mathcal{B}$;
in power electronics (see, e.g., \cite{KarGK16}) and Global Position System (GPS) (see, e.g., \cite{Teu98a}),
$\hbx$ is a deterministic vector in a infinite box, i.e., $\mathcal{B}=\mathbb{Z}^n$.

One of the most commonly used methods to detect $\hbx$ is to solve the following
 Box-constrained Integer Least Squares (BILS) problem:
\beq
\label{e:BILS}
\min_{\x\in\mathcal{B}}\|\y-\A\x\|_2^2,
\eeq
whose solution, denoted by $\x^\sBL$, is the maximum likelihood detector of $\hbx$ due to the fact that $\v\sim \mathcal{N}(\boldsymbol{0},\sigma^2 \I)$.

If $\mathcal{B}=\mathbb{Z}^n$ in \eqref{e:box}, then
\eqref{e:model} is referred to as an ordinary linear model,
and the maximum likelihood estimator of $\hbx$, denoted by $\x^\sOL$, is the solution of
the Ordinary Integer Least Squares (OILS) problem:
\beq
\label{e:OILS}
\min_{\x\in\mathbb{Z}^n}\|\y-\A\x\|_2^2.
\eeq

One of the widely used approaches to solving \eqref{e:BILS} or \eqref{e:OILS} is sphere decoding,
which consists of two steps: reduction and discrete search.
Reduction is the process of using some lattice reduction strategies to preprocess
\eqref{e:BILS} or \eqref{e:OILS}.
Discrete search is the process of finding the solution of the preprocessed problem
with certain search algorithm.
The most popular reduction method  for preprocessing the OILS problem \eqref{e:OILS} is the Lenstra-Lenstra-Lov\'asz (LLL)
reduction \cite{LenLL82} \cite{LyuL17}. It consists of two kinds of strategies:
size reductions and column permutations, which reduce the magnitudes of the off-diagonal entries
and reorder the columns of the triangular factor of the QR factorization of $\A$, respectively.
For the BILS problem \eqref{e:BILS}, the size reductions make the constraint box too complicated
to be handled, thus instead of using the LLL reduction, some column reordering strategies
are frequently utilized to preprocess \eqref{e:BILS}.
The commonly used column reordering strategies includes
Vertical-Bell Laboratories Layered Space Time (V-BLAST) \cite{FosGVW99} and
Sorted QR Decomposition (SQRD) \cite{WubBRKK01} which use the information of $\A$ only,
and those developed in \cite{ChaH08,BreC11},
which use not only  the information of $\A$,
but also the information of $\y$ and $\mathcal{B}$.
The most widely used discrete search strategy for \eqref{e:OILS} is the
Schnorr-Euchner search algorithm \cite{SchE94}, which is an improvement of
the Fincke-Pohst search algorithm \cite{FinP85}.
By taking the box-constraint into account, the Schnorr-Euchner search algorithm
has been modified to solve \eqref{e:BILS}, for more details, see,
e.g., \cite{DamGC03} and \cite{ChaH08}.

Although the reduction step can usually reduce the computational cost
of solving \eqref{e:BILS} or \eqref{e:OILS} (see, e.g., \cite{AgrEVZ02,DamGC03, ChaH08, ChaWX13}),
it has been respectively shown in  \cite{Ver89} and
\cite{Boa81} (a simplified proof can be found in \cite{Mic01})
that \eqref{e:BILS} and \eqref{e:OILS} are NP-hard.
Therefore, in practical applications, especially for real-time applications,
an efficient and effective suboptimal algorithm is often used to detect $\hbx$
instead of exactly solving \eqref{e:BILS} or \eqref{e:OILS} to get the optimal solution.

For the OILS problem, the ordinary rounding detector $\x^\sOR$
and the Babai detector $\x^\sOB$,
which are respectively obtained by the Babai rounding off and nearest plane algorithms \cite{Bab86},
are frequently used as suboptimal detectors for $\hbx$.
(In the ordinary case, actually it is more appropriate to use the term ``estimator'' than ``detector''.
However, as we mainly focus on the box constrained case in this paper
and ``detector'' is the standard term used in communications in this case.
For simplicity, we will use ``detector'' for both cases.)
By taking the box constraint \eqref{e:box} into account, one can easily modify the algorithms for
$\x^\sOR$ and $\x^\sOB$ to get box-constrained rounding detector  $\x^\sBR$
and box-constrained Babai detector  $\x^\sBB$ for $\hbx$ satisfying
both \eqref{e:model} and \eqref{e:box}.
In communications, $\x^\sBR$ is referred to as zero-forcing detector,
while $\x^\sBB$  is called the nulling and canceling detector \cite{FosGVW99}
(also called zero-forcing decision-feed-back equalization detector \cite{DamGC03}).

One of the most popular measures to characterize how good a detector is its success probability,
which is the probability of the detector being equal to $\hbx$,
see e.g., \cite{HasB98, ChaWX13,WenTB16, WenC17}.
There are some other measures to characterize the performance of a detector, such as the bit error rate (see, e.g., \cite{QianM17}).

For the estimation of $\hbx$ in the ordinary linear model \eqref{e:model},
the formulas of the success probability $P^\sOR$ of the rounding detector $\x^\sOR$
and the success probability $P^\sOB$ of the Babai detector  $\x^\sOB$
have been given in \cite{WenTB16} and \cite{ChaWX13}, respectively.
Equivalent formulas of $P^\sOR$ and $P^\sOB$ were given earlier in \cite{Teu98a},
which considers the OILS problem in a different format in GPS.
It is shown in \cite{Teu98a} that  $P^\sOR\leq P^\sOB$.
The success probability $P^\sOL$ of the solution $\x^\sOL$ of \eqref{e:OILS} was
given in \cite{Teu99} which considers the OILS problem in a different format
in GPS. Furthermore, it has been shown in \cite{Teu99}
that $\x^\sOL$ has the highest success probability among a class of the so-called admissible estimators,
which include $\x^\sOR$ and $\x^\sOB$.

For the detection of $\hbx$ satisfying both \eqref{e:model} and \eqref{e:box},
it is also generally believed that
the success probability $P^\sBR$ of the rounding detector $\x^\sBR$ is not larger than
the success probability $P^\sBB$ of the Babai detector  $\x^\sBB$,
and $P^\sBB$ is not larger than the success probability $P^\sBL$
of the solution $\x^\sBL$ of \eqref{e:BILS}.

Since detecting $\hbx$, which can be deterministic or random in the box $\mathcal{B}$,
from \eqref{e:model} and \eqref{e:box} arises from many applications,
and success probability is one of the often used measures of the goodness of a detector,
this paper develops formulas for the success probabilities of the rounding, Babai and maximal likelihood detectors
and investigate their relationships.
Specifically, the contributions of this paper are summarized as follows
(part of this work has been presented in a conference paper \cite{WenCT17a}):

\begin{enumerate}
\item We present formulas for the success probabilities $P_\sD^\sBR$ and $P_\sR^\sBR$ of
the box-constrained rounding detector $\x^\sBR$ in  Theorems \ref{t:PBRD} and \ref{t:PBRR},
corresponding to the case that $\hbx$ is a deterministic parameter vector and  the case that $\hbx$
is uniformly distributed over $\mathcal{B}$ (an assumption often made for multi-input multi-output (MIMO) applications,
see, e.g., \cite{JalO05}), respectively.
We also give a formula for the success probability $P_\sD^\sBB$ of the box-constrained
Babai  detector $\x^\sBB$ for the case that $\hbx$ is  deterministic in  Theorem \ref{t:PBBD},
and develop formulas for the success probabilities
$P_\sD^\sBL$ and $P_\sR^\sBL$ of the BILS detector $\x^\sBL$ (i.e., the solution to \eqref{e:BILS}) in Theorems \ref{t:PBLD} and \ref{t:PBLR},
corresponding to the case that $\hbx$ is deterministic and
$\hbx$ is uniformly distributed over $\mathcal{B}$, respectively.

\item Since it is difficult to compute $P_\sD^\sBR$, $P_\sR^\sBR$ and $P_\sD^\sBB$,
we present some good bounds on them in Corollaries \ref{c:boundPBRD}-\ref{c:PBBDLUB}, respectively,
and these bounds can easily be computed.
For the same reason, we present good upper bounds on $P_\sD^\sBL$ and $P_\sR^\sBL$
in Theorems \ref{t:PBLDUB} and \ref{t:PBLRUB}, respectively.

\item
We give relations among various success probabilities -- the major contribution of this paper.
We compare  the success probabilities of the same type of detectors in different circumstances
in Corollaries \ref{c:boundPBRD}-\ref{c:POLBLR} and Theorem \ref{t:PBBR}.
We also compare the success probabilities of the three  types of detectors in the same circumstances.
Specifically, Example \ref{ex:rounding}
shows that $P_\sD^\sBR$ can be larger than $P_\sD^\sBB$,
while  Theorem \ref{t:PBRPBB} rigorously shows that $P_\sR^\sBR\leq P_\sR^\sBB$;
and Example \ref{ex:BBBL} shows that $P_\sD^\sBB$ can be larger than $P_\sD^\sBL$,
although it is true that $P_\sR^\sBB\leq P_\sR^\sBL$ (this result can be obtained from, e.g., \cite[p18]{Kuh06}).
\end{enumerate}

The practical significance of our contribution is as follows.
Firstly, as can be seen from Sec. \ref{ss:BRBB}, the complexities of computing $\x^\sBR$
and $\x^\sBB$ are the same and are dominated by the QR factorization of $\A$,
thus, the proved inequality $P_\sR^\sBR\leq P_\sR^\sBB$  indicates that in practical applications,
one should usually use $\x^\sBB$ instead of $\x^\sBR$ to detect $\hbx$
if $\hbx$ is uniformly distributed over the constraint box $\mathcal{B}$.
Secondly, since $P_\sR^\sBB\leq P_\sR^\sBL$, if $P_\sR^\sBB$, which can be efficiently computed
(see Theorem \ref{t:PBBR}), is close to 1, then
one can just use $\x^\sBB$ to detect $\hbx$, and hence there is no need to spend extra time to obtain $\x^\sBL$.
Thirdly,  from simulations in Sec.\ \ref{s:sim}, we can see that the upper bound on
$P_\sR^\sBL$ given in Theorem \ref{t:PBLRUB} is close to $P_\sR^\sBL$.
Thus the former can be used as an approximation to the latter, which is difficult to compute.
Fourthly, it is known that $\x^\sBL$ has the highest success probability over all the detectors
(see, e.g., \cite[P.18]{Kuh06}) when $\hbx$ is uniformly distributed
over the constraint box $\mathcal{B}$.
Thus, if the upper bound on $P_\sR^\sBL$ is much smaller than 1,
then there is no detector which can  detect $\hbx$ with high probability
and one should try to improve the physical setting.

The rest of the paper is organized as follows.
We present formulas for $P_\sD^\sBR$, $P_\sR^\sBR$, $P_\sD^\sBB$,
$P_\sR^\sBB$, $P_\sD^\sBL$ and $P_\sR^\sBL$ in Section \ref{s:PBRBB}.
In Section \ref{s:PRPBrelation}, we study the relationships among them.
Simulation tests to illustrate our main results are provided in Section \ref{s:sim}.
Finally, this paper is summarized in Section \ref{s:sum}.

{\bf Notation}.
We use $\e_i$ to denote the $i$-th column of the identity matrix $\I$.
For $\x\in \mathbb{R}^n$, we use $\lfloor \x\rceil$ to denote its nearest integer vector, i.e.,
each entry of $\x$ is rounded to its nearest integer (if there is a tie, the smaller integer is chosen).
For a vector $\x$, $\x_{i:j}$ denotes the subvector of $\x$ formed by entries $i, i+1, \ldots,j$.
For a matrix $\A$, $\A_{i:j,i:j}$ denotes the submatrix of $\A$ formed by rows and columns $i, i+1, \ldots,j$.
For a box $\mathcal{B}= \{\x \in \mathbb{Z}^{n}:\ell \leq \x\leq \u,\; \ell, \u\in \mathbb{Z}^{n}\}$, sometimes we also write it as
$\mathcal{B}=\prod_{i=1}^n {\cal B}_i$ with ${\cal B}_i= \{x_i \in \mathbb{Z}: \ell_i \leq x_i \leq u_i,\; \ell_i, u_i \in \mathbb{Z}\}$.
For a random vector $\v$ following the normal distribution with mean $\bar{\v}$ and covariance matrix $\bSigma$,
we write $\v \sim {\cal N}(\bar{\v}, \bSigma)$.

For the sake of reading convenience, we provide a list of success probability symbols
and the corresponding detectors in Table \ref{t:symbol}.
Note that this paper is mainly concerned with  the quantities in the second part of this table,
although the quantities in the first part are also involved in some results.
For the ordinary case, $\hbx$ is a fixed unknown integer vector, so the three quantities
in the first part of the table do not need to have the  subscript D or R.
\begin{table}[h]
\begin{center}
\caption{Symbols for success probabilities of some detectors} \label{t:symbol}
\begin{tabular}{|c|c|} \hline
Symbol & Detector \\ \hline
$P^\sOR$  &  ordinary rounding     \\
$P^\sOB$ &  ordinary Babai    \\
$P^\sOL$  &   solution for ordinary Integer Least Squares (ILS)  \\ \hline
$P_\sD^\sBR$  & deterministic box-constrained rounding    \\
$P_\sR^\sBR$  & random box-constrained rounding   \\
$P_\sD^\sBB$  & deterministic box-constrained Babai  \\
$P_\sR^\sBB$  & random box-constrained Babai  \\
$P_\sD^\sBL$  & deterministic box-constrained ILS  \\
$P_\sR^\sBL$  & solution for random box-constrained ILS \\  \hline
\end{tabular}
\end{center}
\end{table}

\section{Success probabilities of some detectors}
\label{s:PBRBB}

In this section, we derive formulas for $P_\sD^\sBR$, $P_\sR^\sBR$, $P_\sD^\sBB$,
$P_\sD^\sBL$ and $P_\sR^\sBL$ (see Table \ref{t:symbol}).
Note that the formula for $P_\sR^\sBB$ has been derived in \cite[Th.1]{WenC17}.

\subsection{Definitions of $\x^\sBR$ and  $\x^\sBB$}
\label{ss:BRBB}

In this subsection, we introduce
the box-constrained rounding detector $\x^\sBR$ and the box-constrained Babai detector $\x^\sBB$.

Let $\A$ in \eqref{e:model} have the following QR factorization
\beq
\A=\Q\R,
\label{e:qr}
\eeq
where $\Q\in \Rmbn$ has orthonormal columns (i.e., $\Q^T\Q=\I$) and $\R\in \Rnbn$ is nonsingular upper triangular with positive diagonal entries.
Define $\tby=\Q^T\y$ and $\tilde{\v} = \Q^T \v$.
Then, left multiplying both sides of \eqref{e:model} by $\Q^T$ yields
\beq
\label{e:modelqr}
\tby=\R\hbx+\tilde{\v}, \quad \tilde{\v} \sim \mathcal{N}(\0, \sigma^2 \I)
\eeq
and the BILS problem \eqref{e:BILS} is equivalent to
\beq
\label{e:BILSR}
\min_{\x\in\mathcal{B}}\|\tby-\R\x\|_2^2.
\eeq
Later on we mainly work on the transformed model \eqref{e:modelqr} and the transformed BILS problem \eqref{e:BILSR}.

Let
\beq
\label{e:d}
\d=\R^{-1}\tby,
\eeq
which is the real solution to \eqref{e:BILSR} or \eqref{e:BILS} with the box ${\cal B}$ replaced by $\Rn$.
The box-constrained rounding detector $\x^\sBR$ of $\hbx$ in \eqref{e:modelqr} is computed as follows
(see, e.g., \cite{Lar09}):
\beq
\label{e:Rounding}
\begin{split}
x_i^\sBR=
\begin{cases}
\ell_i, \quad \mbox{ if } \lfloor d_i\rceil< \ell_i,\\
\lfloor d_i\rceil, \mbox{ if }  \ell_i  \leq \lfloor d_i\rceil \leq u_i,\quad i=1, \ldots, n. \\
u_i,\quad \mbox{ if } \lfloor d_i\rceil >u_i,
\end{cases}
\end{split}
\eeq
The box-constrained Babai detector $\x^\sBB$ is computed in the following way (see, e.g., \cite{Lar09}):
\beq \label{e:Babai}
\begin{split}
& c_{i}=(\ty_{i}-\sum_{j=i+1}^nr_{ij}x_j^\sBB)/r_{ii}, \mbox{ with } \sum_{j=n+1}^n r_{nj}x_j^\sBB =0, \\
& x_i^\sBB=
\begin{cases}
\ell_i, & \mbox{ if }\   \lfloor c_i\rceil< \ell_i,\\
\lfloor c_i\rceil, & \mbox{ if }\    \ell_i  \leq \lfloor c_i\rceil \leq  u_i,\,\; i=n,  \ldots, 1.\\
u_i, & \mbox{ if }\    \lfloor c_i\rceil > u_i,
\end{cases}
\end{split}
\eeq

\subsection{Success probability of the box-constrained rounding detector}
\label{ss:PBR}

In this subsection, we develop formulas for $P_\sD^\sBR$ and $P_\sR^\sBR$,
which are the success probabilities of $\x^\sBR$ for
deterministic $\hbx$ and for random $\hbx$ which is uniformly distributed over $\mathcal{B}$,  respectively.
We also give a lower bound on them.
We first present a formula for $P_\sD^\sBR$.

\begin{theorem}
\label{t:PBRD}
Let $\hbx$ in \eqref{e:model} or \eqref{e:modelqr} be a deterministic integer parameter vector that satisfies \eqref{e:box}, then
\beq
\label{e:PBRD}
P_\sD^\sBR=\frac{\det(\R)}{(\sqrt{2\pi} \sigma)^{n}}\int_{{\cal I}_n}\cdots \int_{{\cal I}_1}
\exp\left(-\frac{\|\R\bxi\|^2_2}{2\sigma^2}\right)d\xi_1\cdots d\xi_n,
\eeq
where
\begin{equation}
\label{e:I}
{\cal I}_i:={\cal I}_i(\hbx) =
\begin{cases}
(-\infty, \frac{1}{2}], \quad\mbox{ if } \hx_i=\ell_i \quad \\
(-\frac{1}{2}, \frac{1}{2}], \quad \, \;\mbox{ if } \ell_i<\hx_i< u_i, \,\; i=1,\ldots, n.\\
(-\frac{1}{2}, \infty), \quad \mbox{ if } \hx_i=u_i
\end{cases}
\end{equation}
\end{theorem}

\begin{IEEEproof}
Since $\hbx$ is deterministic and $\tilde{\v}\sim \mathcal{N}(\0,\sigma^2 \I)$,
by \eqref{e:modelqr} and \eqref{e:d}, we have
\beq
\label{e:ddis}
\d -\hbx =\R^{-1}\tilde{\v}\sim \mathcal{N}(\0,\sigma^2(\R^T\R)^{-1}).
\eeq
Then, by  \eqref{e:Rounding} and \eqref{e:I}, we can conclude that
\beq
\label{e:BRcond}
\x^\sBR=\hbx\Longleftrightarrow  \e_i^T\R^{-1}\tilde{\v}  \in {\cal I}_i, \ \ i=1,\ldots, n.
\eeq
Therefore, \eqref{e:PBRD} holds.
\end{IEEEproof}

From \eqref{e:PBRD} and \eqref{e:I}, we see that  $P_\sD^\sBR$  depends on the position of $\hbx$ in the box $\mathcal{B}$,
thus we also write $P_\sD^\sBR$  as $P_\sD^\sBR(\hbx)$.
To compute $P_\sD^\sBR$, we need to know the positions of $\hat{x}_i$ on $[\ell_i, u_i]$
for $i=1,\ldots, n$. In practice this information is  unknown.
However,  it is easy to observe from \eqref{e:PBRD} that $P_\sD^\sBR$ has a lower bound
which does not rely on this information.
\begin{corollary}
\label{c:boundPBRD}
Under the conditions of Theorem \ref{t:PBRD}, we have
\begin{align*}
P_\sD^\sBR &\geq  \frac{\det(\R)}{(\sqrt{2\pi } \sigma)^{n}}
\int_{-1/2}^{1/2}\cdots \int_{-1/2}^{1/2}
  \exp\left(-\frac{\|\R\bxi\|^2_2}{2\sigma^2}\right)d\xi_1\cdots d\xi_n \\
&= P^\sOR,
\end{align*}
where the lower bound is reached if and only if $\ell_i < \hx_i <u_i$ for $i=1,\ldots,n$.
\end{corollary}

The lower bound on $P_\sD^\sBR$  in Corollary \ref{c:boundPBRD} is equal to the success probability of the
ordinary rounding detector $P^\sOR$, which can be found in \cite[Th.\ 1]{WenTB16}.
It is easy to understand this.
In fact, the ordinary case can be regarded as a special case of the box-constrained case
with $\ell_i = -\infty$ and $u_i=\infty$, thus, $\ell_i < \hx_i <u_i$  for $i=1,\ldots,n$.

The following theorem gives a formula for $P_\sR^\sBR$.
\begin{theorem}
 \label{t:PBRR}
Suppose that $\hbx$ in \eqref{e:model} is a random integer parameter vector that is
uniformly distributed over $\mathcal{B}$, and $\hbx$ and $\v$ are independent, then
\begin{align}
P_\sR^\sBR
  = &\   \frac{\det(\R)}{(\sqrt{2\pi} \sigma)^{n}}
\left(\alpha_n  \int_{-\infty}^{\infty}d \xi_n +\beta_n  \int_{-1/2}^{1/2} d \xi_n\right) \nonumber \\
&\ \times  \cdots
 \left( \alpha_1 \int_{-\infty}^{\infty} d \xi_1 +\beta_1  \int_{-1/2}^{1/2} d \xi_1  \right)
  \exp\left(-\frac{\|\R\bxi\|^2_2}{2\sigma^2}\right)  \label{e:PBRR1}  \\
=& \   \frac{\det(\R)}{(\sqrt{2\pi} \sigma)^{n}}
\sum_{\begin{subarray}{l}\omega_i\in\{\alpha_i, \beta_i\} \\ i=1,\cdots, n\end{subarray}}
\biggl(\prod_{i=1}^n\omega_i \biggr)\int_{\mathcal{\bar{I}}(\omega_n)}\cdots\int_{\mathcal{\bar{I}}(\omega_1)} \nonumber \\
&\exp\left(-\frac{\|\R\bxi\|^2_2}{2\sigma^2}\right)d \xi_1 \cdots  d \xi_n,
\label{e:PBRR11}
\end{align}
where for $1\leq i\leq n$,
\beq
\label{e:alphabeta}
\alpha_i =  \frac{1}{u_i-\ell_i+1}, \ \ \beta_i = \frac{u_i-\ell_i}{u_i-\ell_i+1},
\eeq
\beqnn
\mathcal{\bar{I}}(\omega_i) =
\begin{cases}
      (-\infty, +\infty)& \mbox{ if } \omega_i=\alpha_i\\
      [-1/2,1/2] & \mbox{ if } \omega_i=\beta_i
   \end{cases}.
\eeqnn
\end{theorem}

\begin{IEEEproof}
See Appendix \ref{ss:pfPBRR}.
\end{IEEEproof}

Although the formula for $P_\sD^\sBR$ given in Theorem \ref{t:PBRR} could be used for computation,
it may be too expensive when the dimension $n$ is a little large,
hence, Theorem \ref{t:PBRR} is of little practical use.

By Theorem \ref{t:PBRR}, we get the following corollary.

\begin{corollary} \label{c:boundPBRR}
Under the  conditions of Theorem \ref{t:PBRR}, we have
\begin{align}
\label{e:PBRRld}
 P_\sR^\sBR > P^\sOR.
\end{align}
and
\beq \label{e:PBRRlim}
\lim_{\scriptsize \begin{array}{c} l_i \rightarrow -\infty \ \mathrm{or}\ u_i \rightarrow \infty \\ i=1,\ldots,n\end{array}} P_\sR^\sBR = P^\sOR.
\eeq
\end{corollary}
\begin{IEEEproof}
From \eqref{e:PBRR1}, by applying Corollary \ref{c:boundPBRD},  one can obtain \eqref{e:PBRRld}.
From \eqref{e:PBRR1} and Corollary \ref{c:boundPBRD}, it is easy to see that \eqref{e:PBRRlim} holds.
\end{IEEEproof}

This corollary shows the relation  between $P_\sR^\sBR$ and $P^\sOR$.
The latter is a strict lower bound on the former for a finite box ${\cal B}$.
But the two quantities will be close when ${\cal B}$ is big enough.

\subsection{Success probability of the box-constrained Babai detector}

In this subsection, we  develop a formula for  $P_\sD^\sBB$.
Since $P_\sD^\sBB$ depends on the position of $\hbx$ in the box $\mathcal{B}$,
we also give a lower bound and an upper bound on $P_\sD^\sBB$.
The following theorem presents a formula for $P_\sD^\sBB$.

\begin{theorem}
\label{t:PBBD}
Let $\hbx$ in \eqref{e:model} be a deterministic integer parameter vector that satisfies \eqref{e:box}, then
\beq
P_\sD^\sBB=\prod_{i=1}^n\omega_i(r_{ii}),
\label{e:PBBD}
\eeq
where
\begin{equation}\label{e:omega}
\omega_i(r_{ii})=
\begin{cases}
\frac{1}{2}\big[1+\phi_\sigma(r_{ii})\big], &
       \mbox{ if }\  \hx_i=\ell_i  \mbox{ or }  \hx_i=u_i\\
\phi_\sigma(r_{ii}), &
 \mbox{ if }\  \ell_i<\hx_i< u_i
\end{cases}
\end{equation}
with
\beq  \label{e:varphi}
\begin{split}
\phi_\sigma(\zeta)
& =\textup{erf}\left(\zeta/(2\sqrt{2}\sigma)\right)
  \!=\! \frac{2}{\sqrt{2\pi}}\int_{0}^{\frac{\zeta}{2\sigma}}\exp\big(\!\!-\! \frac{1}{2}t^2\big)dt  \\
& = \frac{\zeta}{\sqrt{2\pi}\sigma} \int_{-\frac{1}{2}}^{\frac{1}{2}} \exp\big(\! -\frac{\zeta^2}{2\sigma^2}t^2\big)dt.
\end{split}
\eeq
\end{theorem}

Theorem \ref{t:PBBD} was originally given in the MSc thesis \cite{Han12},
supervised by the second author of this paper.
It can be proved easily by following the proof of \cite[Th. 1]{WenC17},
so for concise, we omit its proof.

As $P_\sD^\sBB$ in \eqref{e:PBBD} depends on the diagonal entries of $\R$,
we sometimes write $P_\sD^\sBB$ as $P_\sD^\sBB(\R).$
From Theorem \ref{t:PBBD} we observe  that like computing $P_\sD^\sBR$,
computing $P_\sD^\sBB$ needs to know the location of $\hat{\x}$ in the box $\mathcal{B}$.
This requirement is not practical.
However, by \eqref{e:PBBD} and \eqref{e:omega}, we can find
a lower bound and an upper bound on $P_\sD^\sBB$,  which do not need priori information on  $\hbx$.

\begin{corollary} \label{c:PBBDLUB}
Under the  conditions of Theorem \ref{t:PBBD},
\begin{equation} \label{e:BBLUBD}
 P^\sOB =\prod_{i=1}^n \phi_\sigma(r_{ii}) \leq P_\sD^\sBB  \leq \frac{1}{2^n} \prod_{i=1}^n (1+\phi_\sigma(r_{ii})),
\end{equation}
where the lower bound is reached if and only if $\ell_i < \hx_i <u_i$ for $i=1,\ldots,n$,
and the upper  bound is reached if and only if $\hx_i =\ell_i$ or $\hx_i=u_i$ for $i=1,\ldots,n$.
\end{corollary}

Note that $P^\sOB$ in \eqref{e:BBLUBD} denotes the success probability of the ordinary Babai detector $\x^\sOB$,
and the equality in \eqref{e:BBLUBD} was given by \cite[eq. (11)]{ChaWX13}.

The following theorem from \cite[Th. 1]{WenC17}, provides a formula for  $P_\sR^\sBB$.

\begin{theorem}
\label{t:PBBR}
Suppose that $\hbx$ in \eqref{e:model} is a random integer parameter vector that is
uniformly distributed over $\mathcal{B}$, and $\hbx$ and $\v$ are independent, then
\beq
P_\sR^\sBB=\prod_{i=1}^n \frac{1+(u_i-\ell_i) \phi_\sigma(r_{ii})}{u_i-\ell_i+1},
\label{e:PBBR}
\eeq
where  $\phi_\sigma(\zeta)$ is defined in \eqref{e:varphi}.
Furthermore, we have
\beqnn
 P_\sR^\sBB > P^\sOB
\eeqnn
and
\beqnn
\lim_{\begin{array}{c} \ell_i \rightarrow -\infty \ \mathrm{ or }\  u_i \rightarrow \infty \\ i=1,\ldots,n\end{array}} P_\sR^\sBB = P^\sOB.
\eeqnn
\end{theorem}

\subsection{Success probability of the BILS detector}

In this subsection, we  give formulas for $P_\sD^\sBL$ and $P_\sR^\sBL$.
Since both $P_\sD^\sBL$ are $P_\sR^\sBL$ are complicated to be computed,
we also give an upper bound on each of them.
We first consider the deterministic situation.

\begin{theorem}
\label{t:PBLD}
Let $\hbx$ in \eqref{e:model} be a deterministic integer parameter vector
that satisfies \eqref{e:box} and $\x^\sBL$ be any solution to the BILS problem \eqref{e:BILSR},
then the success probability $P_\sD^\sBL$ of $\x^\sBL$ satisfies
\begin{align}
\label{e:PBLD}
P_\sD^\sBL
=\frac{1}{(\sqrt{2\pi } \sigma)^{n}}\int_{{\cal S}^\sBL}\exp\big(-\frac{1}{2\sigma^2}\|\bxi\|_2^2)d\bxi,
\end{align}
where
\beq
\label{e:S}
{\cal S}^\sBL \!=\!\{\bxi\mid  2(\x-\hbx)^T\R^T\bxi \!\leq \!  \|\R(\x-\hbx)\|_2^2 \  \mbox{ for }\forall \x\in\mathcal{B}\}.
\eeq
\end{theorem}

\begin{IEEEproof}
See Appendix \ref{ss:pfPBLD}.
\end{IEEEproof}

From \eqref{e:S} we see that $P_\sD^\sBL$ depends on $\hbx$, thus we also write
$P_\sD^\sBL$ as $P_\sD^\sBL (\hbx)$.
Like $P_\sD^\sBR$ (see \eqref{e:PBRD}) and $P_\sD^\sBB$ (see \eqref{e:PBBD}),
one  cannot use \eqref{e:PBLD} to compute $P_\sD^\sBL$, since we do not know $\hbx$.
Furthermore, the set ${\cal S}^\sBL$ in  \eqref{e:S} is complicated and it is difficult to calculate
$P_\sD^\sBL$ from the computational perspective, even if $\hbx$ is known.

By Theorem \ref{t:PBLD}, we can obtain the following corollary.
\begin{corollary} \label{c:POLBL}
The success probability $P^\sOL$ of the solution $\x^\sOL$ to the OILS problem,
and $P_\sD^\sBL$ of the solution $\x^\sBL$ to the BILS problem \eqref{e:BILSR} satisfy
\[
P^\sOL\leq P_\sD^\sBL.
\]
\end{corollary}

{\bf Proof}.
By using the same method as that used for deriving $P_\sD^\sBL$ in Theorem \ref{t:PBLD}, one can easily obtain
\[
P^\sOL
=\frac{1}{(\sqrt{2\pi } \sigma)^{n}}
\int_{\mathcal{S}^{\sOL}}\exp\big(-\frac{1}{2\sigma^2}\|\bxi\|_2^2)d\bxi,
\]
where
\[
\mathcal{S}^{\sOL}\!=\!\{\bxi\mid  2(\x-\hbx)^T\R^T\bxi \!\leq \!  \|\R(\x-\hbx)\|_2^2 \mbox{ for }\,\forall \, \x\in \mathbb{Z}^n\}.
\]
Note that $\mathcal{S}^{\sOL}$ is the same as $\mathcal{S}^{\sBL}$ in \eqref{e:S}, except that the constraint set
${\cal B}$ is replaced by $\Zn$.
Thus, $\mathcal{S}^{\sOL}\subseteq \mathcal{S}^\sBL$.
Then,  comparing the expressions for $ P^\sOL$ here and $P_\sD^\sBL$ in \eqref{e:PBLD}, we obtain $ P^\sOL\leq P_\sD^\sBL$.
\ \ $\Box$

In the following, we give some upper bounds on $P_\sD^\sBL (\hbx)$,
one of them can be calculated easily without using any information of $\hbx$.

\begin{theorem} \label{t:PBLDUB}
Let $\hbx$ in \eqref{e:model} be a deterministic integer parameter vector
that satisfies \eqref{e:box} and $\x^\sBL$ be any solution to the BILS problem
\eqref{e:BILSR}.
Let $\phi_\sigma(\cdot)$ be defined in \eqref{e:varphi}.
For any $\x$ such that $\x\in {\cal B}$ and $\x\neq \hbx$,
\begin{align}
\label{e:PBLDUB1}
P_\sD^\sBL\leq
\frac{1}{2}\big[1+\phi_\sigma( \|\R(\x-\hbx)\|_2)\big].
\end{align}
In particular,
\begin{align}
\label{e:PBLDUB}
P_\sD^\sBL\leq \frac{1}{2}\big[1+\phi_\sigma\big(\min\limits_{1\leq i\leq n}\|\R_{1:i,i}\|_2\big)\big].
\end{align}

\end{theorem}

\begin{IEEEproof}
See Appendix \ref{ss:pfPBLDUB}.
\end{IEEEproof}

Note that the upper bound given by \eqref{e:PBLDUB} is independent of the
box $\mathcal{B}$, thus it also holds when $u_i-\ell_i, 1\leq i\leq n,$ tends to infinity.
Hence, the right-hand side of \eqref{e:PBLDUB} is also an upper bound on $P^\sOL$.
Furthermore, from Theorem \ref{t:PBLD}, we can see that $P_\sD^\sBL$ becomes larger as the
box gets smaller for fixed $\A$, so the upper bound \eqref{e:PBLDUB} becomes
sharper as the box gets smaller.

In the rest of this section, we consider the case that  $\hbx$ is uniformly distributed over $\mathcal{B}$.
By using the technique for deriving  \eqref{e:PBRR} in Appendix \ref{ss:pfPBRR}, we can easily obtain the following result.

\begin{theorem} \label{t:PBLR}
Suppose that $\hbx$ in \eqref{e:model} is a random integer parameter vector that is
uniformly distributed over $\mathcal{B}$, and $\hbx$ and $\v$ are independent.
Let  $\x^\sBL$ be any   solution to the BILS problem \eqref{e:BILSR},
then the success probability $P_\sR^\sBL$ of $\x^\sBL$ satisfies
\begin{align}
\label{e:PBLR}
P_\sR^\sBL=\frac{1}{\prod_{i=1}^n(u_i-\ell_i+1)}
\sum_{\forall \bar{\x}\in\mathcal{B}}P_\sD^\sBL (\bar{\x}),
\end{align}
where $P_\sD^\sBL (\bar{\x})$ denotes the success probability of the BILS estimator
when $\hbx=\bar{\x}$.
\end{theorem}

Although in theory it is possible to obtain $P_\sR^\sBL$,
it is challenging to compute it numerically for high dimension $n$ and big box ${\cal B}$.
But the formula \eqref{e:PBLR} is useful in analysis.

By Theorem \ref{t:PBLR} and Corollary \ref{c:POLBL}, we can easily obtain the following result.
\begin{corollary} \label{c:POLBLR}
The success probability $P^\sOL$ of the solution $\x^\sOL$ to the OILS problem,
and $P_\sR^\sBL$ of the solution $\x^\sBL$ to the BILS problem \eqref{e:BILSR} satisfy
\[
 P^\sOL\leq P_\sR^\sBL.
\]
\end{corollary}

By \eqref{e:PBLDUB} and \eqref{e:PBLR}, one can easily see that the right-hand
side of \eqref{e:PBLDUB} is also an upper bound on $P_\sR^\sBL$.
But we can get a  sharper upper bound.
Before presenting the upper bound, we need to introduce the following lemma.

\begin{lemma}
\label{l:integralineq1}
Suppose that $\U\in \Rnbn$ is an upper triangular matrix with positive diagonal entries,
and $a_i>0$ (where either $a_i <\infty$ or $a_i=\infty$)  for $1\leq i \leq n$.
Then
\beq \label{e:integralineq1}
\begin{split}
&\int_{-a_n}^{a_n}\cdots \int_{-a_1}^{a_1}
\exp\left(- \|\U\bxi\|^2_2 \right)d\xi_1\cdots d\xi_n \nonumber\\
\leq  & \prod_{i=1}^n \int_{-a_i}^{a_i}\exp\left(- u_{ii}^2t^2 \right)dt.
\end{split}
\eeq
\end{lemma}

\begin{IEEEproof}
See Appendix \ref{ss:pfintegralineq1}.
\end{IEEEproof}

Here we make a remark.
If $a_i=1/2$ for $i=1,\ldots,n$ in \eqref{e:integralineq1}, an inequality  equivalent to \eqref{e:integralineq1}
was derived and used to show
that  the success probability of ordinary rounding detectors cannot be larger than
the success probability of ordinary Babai detectors in \cite{Teu98a}.

The following theorem gives an upper bound on $P_\sR^\sBL$, which can easily be computed.

\begin{theorem}
\label{t:PBLRUB}
Suppose that $\hbx$ in \eqref{e:model} is a random integer parameter vector that is
uniformly distributed over $\mathcal{B}$, and $\hbx$ and $\v$ are independent.
Let  $\x^\sBL$ be any   solution to the BILS problem \eqref{e:BILSR},
then the success probability $P_\sR^\sBL$ of $\x^\sBL$ satisfies
\begin{align}
\label{e:PBLRUB}
P_\sR^\sBL\leq
&\prod_{i=1}^n \biggl( \frac{1}{u_i-\ell_i+1}
+ \frac{u_i-\ell_i}{u_i-\ell_i+1} \phi_\sigma\biggl(\frac{\|\R\e_i\|_2^2}{r_{ii}}\biggr)\biggr) \nonumber\\
:= &\mu^\sBL.
\end{align}
\end{theorem}

\begin{IEEEproof}
See Appendix \ref{ss:pfPBLRUB}.
\end{IEEEproof}

Here we give a remark about a practical use of the upper bound $\mu^\sBL$.
If  $\mu^\sBL$ is much smaller than 1, then one may give up detection
 without  bothering to find a detector.
Later in Section \ref{s:PBBBL} we will give another remark about $\mu^\sBL$
when we compare $P_\sR^\sBL$ and $P_\sR^\sBB$.
In Section \ref{s:sim} we will give some numerical examples to show how tight the upper bound is.

\section{Relationships among $P^\sBR$, $P^\sBB$ and $P^\sBL$}
\label{s:PRPBrelation}

In this section, we investigate the relationships among $P^\sBR$, $P^\sBB$ and $P^\sBL$.
Specifically, on the one hand, we give a simple example to show that $P_\sD^\sBR> P_\sD^\sBB$
and then rigorously show that $P_\sR^\sBB> P_\sR^\sBR$.
On the other hand, since it is well-known that $P_\sR^\sBB\leq P_\sR^\sBL$ (see \cite[P.18]{Kuh06}),
we give a simple example to show that $P_\sD^\sBB> P_\sD^\sBL$ may hold.

\subsection{Relationship between $P^\sBR$ and $P^\sBB$}

It has been shown in \cite[eq. (20)]{Teu98a} that the success probability of the ordinary
rounding detector cannot be larger than that of the ordinary Babai detector,
i.e., $P^\sOR \leq P^\sOB$.
For the box-constrained case, if the deterministic $\hbx$ satisfies $\ell_i<x_i<u_i$
for $1\leq i \leq n$, then by Corollaries \ref{c:boundPBRD} and \ref{c:PBBDLUB},
$P_\sD^\sBR=P^\sOR$ and $P_\sD^\sBB=P^\sOB$ which imply that $P_\sD^\sBR \leq P_\sD^\sBB$.
When $x_i=\ell_i$ or $x_i=u_i$ for some $i$, our simulations indicate that
in general the  experimenta success probability of $\x^\sBR$ is smaller than that of $\x^\sBB$. 
However, the following example shows that in this case
it is possible that $P_\sD^\sBR>P_\sD^\sBB$.

\begin{example}\label{ex:rounding}
Suppose that in \eqref{e:modelqr} $\sigma=1$, $\R=\bmx 2& -1\\ 0 & 1\emx$,  $\hx_{1}=\ell_{1}$ and
$\hx_{2}=\ell_{2}$. Then, by Theorems \ref{t:PBRD} and \ref{t:PBBD}, we have
\begin{align*}
& P_\sD^\sBR=\frac{2}{2\pi}\int_{-\infty}^{1/2}\int_{-\infty}^{1/2}\exp(-\frac{1}{2}\|\R\bxi\|^2_2)d\xi_2d\xi_1=0.6192, & \\
& P_\sD^\sBB=\frac{1}{4}(1+\phi_1(1))(1+\phi_1(2))=0.5818. &
\end{align*}
Thus, $P_\sD^\sBR>P_\sD^\sBB$.
\end{example}

Unlike the deterministic situation,  when $\hbx$ is uniformly distributed over $\mathcal{B}$,
we will show in Theorem \ref{t:PBRPBB} below that $P_\sR^\sBR \leq P_\sR^\sBB$.

\begin{theorem}
 \label{t:PBRPBB}
Suppose that $\hbx$ in \eqref{e:model} is a random integer parameter vector that is
uniformly distributed over $\mathcal{B}$, and $\hbx$ and $\v$ are independent, then
\beq
\label{e:PBRPBB}
P_\sR^\sBR \leq P_\sR^\sBB.
\eeq
\end{theorem}

\begin{IEEEproof}
See Appendix \ref{ss:PBRPBB}.
\end{IEEEproof}

\subsection{Relationships between $P^\sBB$ and $P^\sBL$} \label{s:PBBBL}

We mentioned before that $\x^\sBL$ is optimal in terms of the success probability if $\hbx$
is uniformly distributed over the constraint box $\mathcal{B}$.
Thus we have
\beq \label{e:PBBBL}
P_\sR^\sBB \leq P_\sR^\sBL.
\eeq
In Theorem \ref{t:PBLRUB} when $\R$ is diagonal,  the upper bound $\mu^\sBL$ on $P_\sR^\sBL$  in \eqref{e:PBLRUB} becomes $P_\sR^\sBB$ (see \eqref{e:PBBR}).
Thus, in this case, \eqref{e:PBBBL} holds with equality.
It is easy to understand this as $\x^\sBB = \x^\sBL$ in this case.
If $\R$ is nearly diagonal, the upper bound $\mu^\sBL$  will be close to  the lower bound $P_\sR^\sBB$,
thus it must be tight.
In practice, if we find $\mu^\sBL$ is close to $P_\sR^\sBB$, then we do not need to solve
the BILS problem to find $\x^\sBL$ and we can just use $\x^\sBB$ as the detector.

Although the inequality \eqref{e:PBBBL} holds,
Example \ref{ex:BBBL} below shows that the success probability $P_\sD^\sBB$ of
the box-constrained Babai detector $\x^\sBB$ can be  larger than
the success probability $P_\sD^\sBL$ of the BILS detector $\x^\sBL$ if $\hbx$ is
a deterministic parameter vector.

\begin{example}
\label{ex:BBBL}
Suppose that in \eqref{e:modelqr}, $\sigma=1$, $\R=\bmx 2 & -1\\ 0 & 2\emx$,
$\mathcal{B} = [1,2]\times [1,2]$ and $\hbx=[ 2,  2]^T$.

Since $\hbx=[ 2,  2]^T$, by  \eqref{e:PBBD}, we have
\begin{align*}
P_\sD^\sBB=\frac{1}{4}[1+\phi_\sigma(2)]^2=0.7079.
\end{align*}
By \eqref{e:S} and some simple calculations, we can see that
$\bxi\in S$ if and only if $\bxi$ satisfies all of the following inequalities:
\begin{align*}
& \xi_2\geq -\frac{1}{2}\xi_1-\frac{5}{4} \ \ (\mbox{take} \;  \x=[1,  1]^T),   \\
&  \xi_1\geq -1\ \  (\mbox{take} \; \x= [ 1,  2]^T),  \\
&  \xi_2\geq \frac{1}{2}\xi_1-\frac{5}{4} \ \ (\mbox{take} \; \x=[ 2, 1]^T).
\end{align*}
Thus,
\begin{align*}
\mathcal{S}^\sBL=&\Big\{\bxi\mid \xi_2\geq \frac{1}{2}\xi_1-\frac{5}{4} , \; \xi_1\geq 0 \Big\}\\
\bigcup &\Big\{\bxi\mid \xi_2\geq -\frac{1}{2}\xi_1-\frac{5}{4}, \;  -1\leq \xi_1\leq 0 \Big\}.
\end{align*}
Then, by  \eqref{e:PBLD}, we obtain that
$$P_\sD^\sBL=\frac{1}{2\pi}\int_{{\cal S}^\sBL}\exp[-\frac{1}{2}(\xi_1^2+\xi_2^2)] d\bxi=0.6845.$$
Thus, $P_\sD^\sBB > P_\sD^\sBL$.
\end{example}

Example \ref{ex:BBBL} shows that $P_\sD^\sBB> P_\sD^\sBL$ may hold when
the true parameter vector $\hbx$ is on the boundary of the constraint box.
But if $\hbx$ is inside the box,
the following theorem shows that $P_\sD^\sBB\leq P_\sD^\sBL$ always holds.

\begin{theorem} \label{t:BBBL1}
Suppose  that $\hbx$ in \eqref{e:model} is a deterministic integer parameter vector which satisfies
$\ell_i< \hat{x}_i< u_i$ for $i=1, \ldots, n$, then
\beq
 \label{e:BBBL1}
P_\sD^\sBB\leq P_\sD^\sBL.
\eeq
\end{theorem}

{\bf Proof}. Since $\ell_i< \hat{x}_i< u_i$ for $i=1, \ldots, n$,
By Corollary \ref{c:PBBDLUB}, $P_\sD^\sBB=P^\sOB$
(recall $P^\sOB$ is the success probability of the ordinary Babai detector).
Let $P^\sOL$ denote the success probability of the solution $\x^\sOL$ to the OILS problem \eqref{e:OILS},
then by \cite{Teu99}, we have $P^\sOB\leq P^\sOL$.
Then, by Corollary \ref{c:POLBL}, one can see that \eqref{e:BBBL1} holds.
\ \ $\Box$

\section{Simulation Results}
\label{s:sim}

In this section, we do numerical tests to illustrate our theoretical findings.
As $\hbx$ is typically assumed to be uniformly distributed over $\mathcal{B}$  in communications,
we consider this case only in this section.
The formulas for $P_\sR^\sBR$, $P_\sR^\sBB$ and $P_\sR^\sBL$ have been derived
in Section \ref{s:PBRBB} and their relationships have been established in Section \ref{s:PRPBrelation}.
We would like to compare them numerically.
However,  the cost of computing $P_\sR^\sBR$ or $P_\sR^\sBL$ is extremely high when $n$ is large,
so we compare the experimental and theoretical success probability of $P_\sR^\sBB$ only.
Note that the experimental success probability of a detector is the number of
correct detection divided by the total number of tests,
and the theoretical success probability of $P_\sR^\sBB$ is obtained by \eqref{e:PBLR}.

Since the column permutation strategy V-BLAST is commonly used in practical applications
to improve the decoding performance of $\x^\sBB$,
we also compute the success probability of $\x^\sBB$ after V-BLAST is applied in computing
the QR factorization of $\A$ to see its effect. This $\x^\sBB$ is referred to as the V-BLAST aided $\x^\sBB$.
Note that $\x^\sBR$ and $\x^\sBL$ are not changed by column permutations.

In the tests, for each fixed $n$, constraint box $\mathcal{B}=[0,u]^n$ and signal-to-noise ratio (SNR), we generated 100 $\A$'s with
$a_{ij}, 1\leq i,j \leq n,$ independently and identically following the standard Gaussian distribution $\mathcal{N}(0,1)$.
Then, for each generated $\A$, we randomly generated 100
$\x$'s$\in \mathbb{Z}^n$ that follow the uniform  distribution over $\mathcal{B}$,
and 100 $\v$'s$\in \mathbb{R}^n$ that follow the  Gaussian distribution $\mathcal{N}(\0,\sigma^2\I)$, where $\sigma$ is found from the following equation (see \cite[Appendix C]{WenWTF18}):
\[
\mbox{SNR}=10\log_{10}\frac{u(u+2)}{12\sigma^2},
\]
and then computed the corresponding vector $\y$ based on the linear model \eqref{e:model}.
For each instance, we  computed $\x^\sBR$ by  \eqref{e:Rounding},
$\x^\sBB$  by \eqref{e:Babai},
the V-BLAST aided $\x^\sBB$ by \eqref{e:Babai},
and  $\x^\sBL$ by using the sphere decoding method in \cite{ChaH08}.
Finally, we computed their {\em experimental} success probabilities,
which are denoted by ``Rounding", ``Babai", ``Babai-VBLAST-E" and ``BILS", respectively.
We also computed  the average of the upper bound $\mu^\sBL$ (after V-BLAST is applied in computing
the QR factorization of $\A$)
on $P_\sR^\sBL$ given in Theorem \ref{t:PBLRUB}, to be denoted by ``BILS-UB",
and computed the average of the {\em theoretical} success probability of the V-BLAST aided
Babai point via \eqref{e:PBBR}, to be denoted  by ``Babai-VBLAST-T".

\begin{figure}[!htbp]
\centering
\includegraphics[width=3.2in]{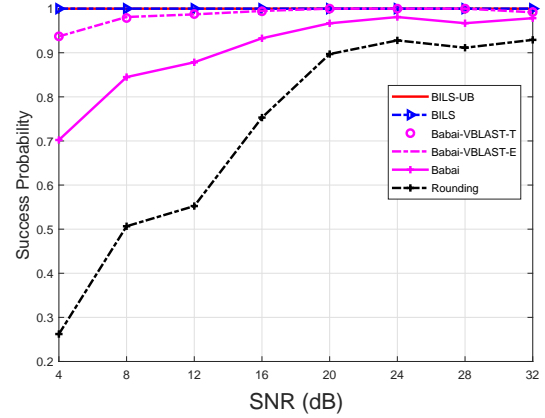}
\caption{Success probabilities versus $\mathrm{SNR}=5:5:40$ dB
for $n=20$ and $\mathcal{B}=[0,1]^n$}
\label{fig:sigma1}
\end{figure}

\begin{figure}[!htbp]
\centering
\includegraphics[width=3.2in]{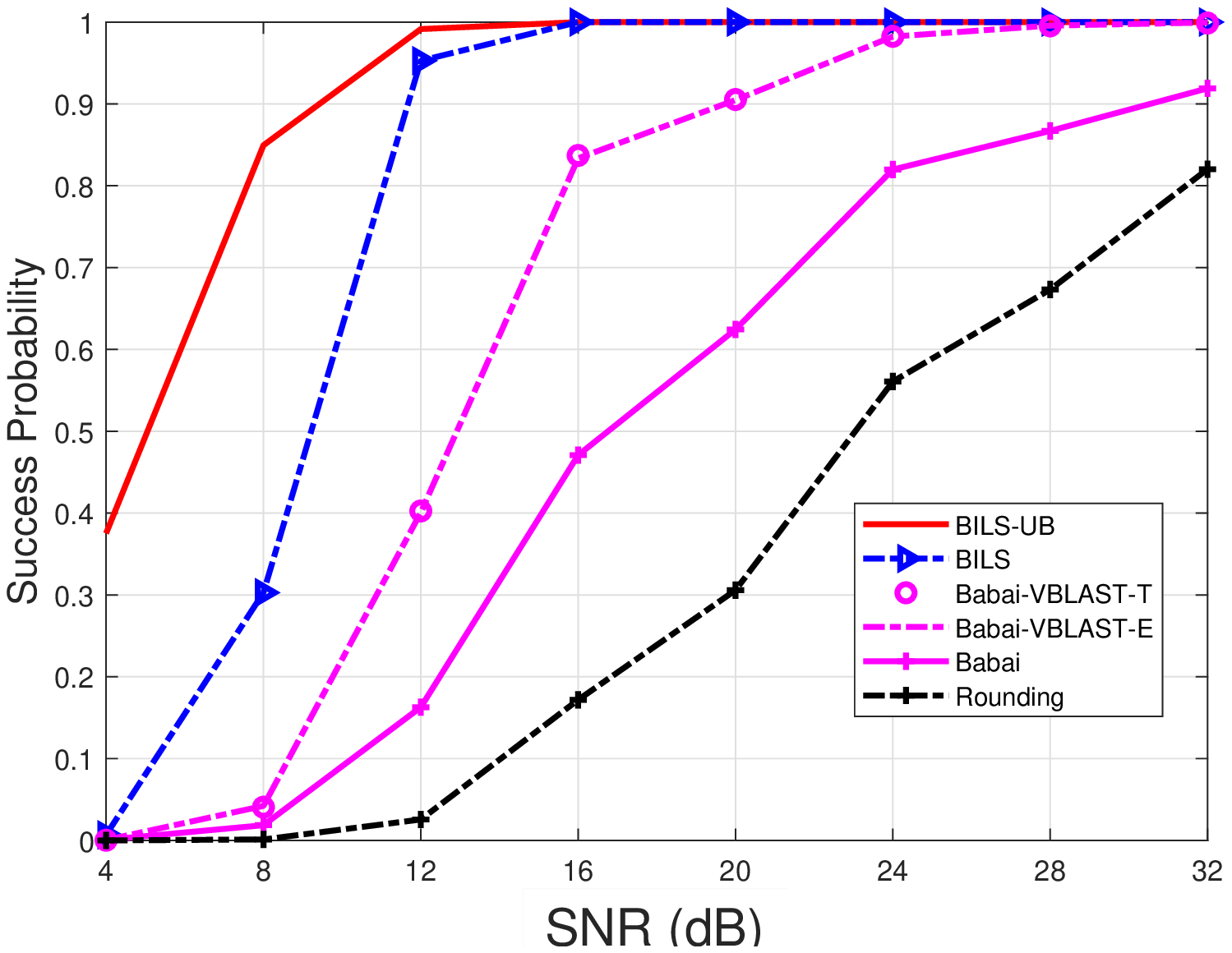}
\caption{Success probabilities versus $\mathrm{SNR}=5:5:40$ dB
for $n=20$ and $\mathcal{B}=[0,7]^n$}
\label{fig:sigma2}
\end{figure}

\begin{figure}[!htbp]
\centering
\includegraphics[width=3.2in]{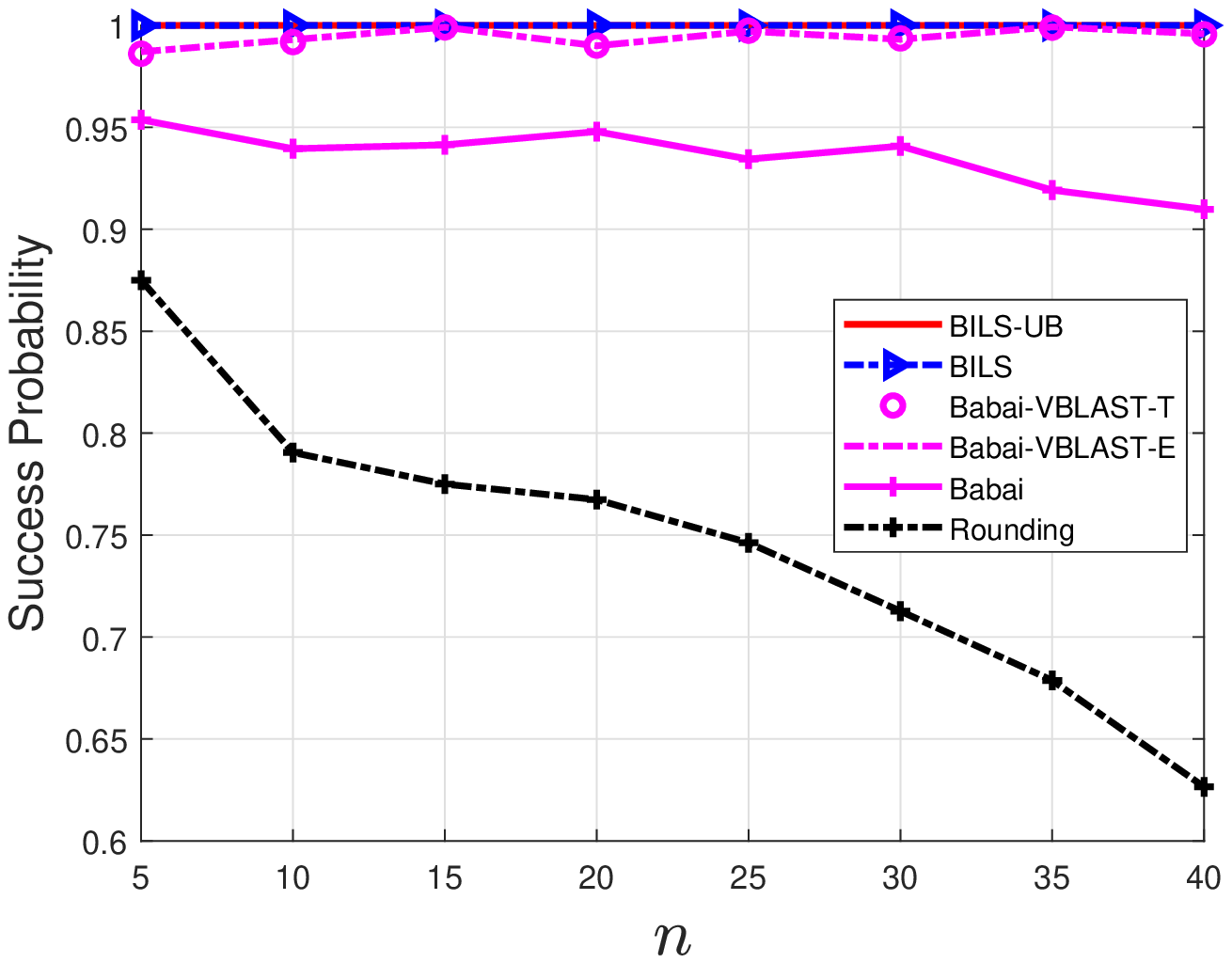}
\caption{Success probabilities versus $n=5:5:40$
for SNR=15 dB and $\mathcal{B}=[0,1]^n$}
\label{fig:n1}
\end{figure}

\begin{figure}[!htbp]
\centering
\includegraphics[width=3.2in]{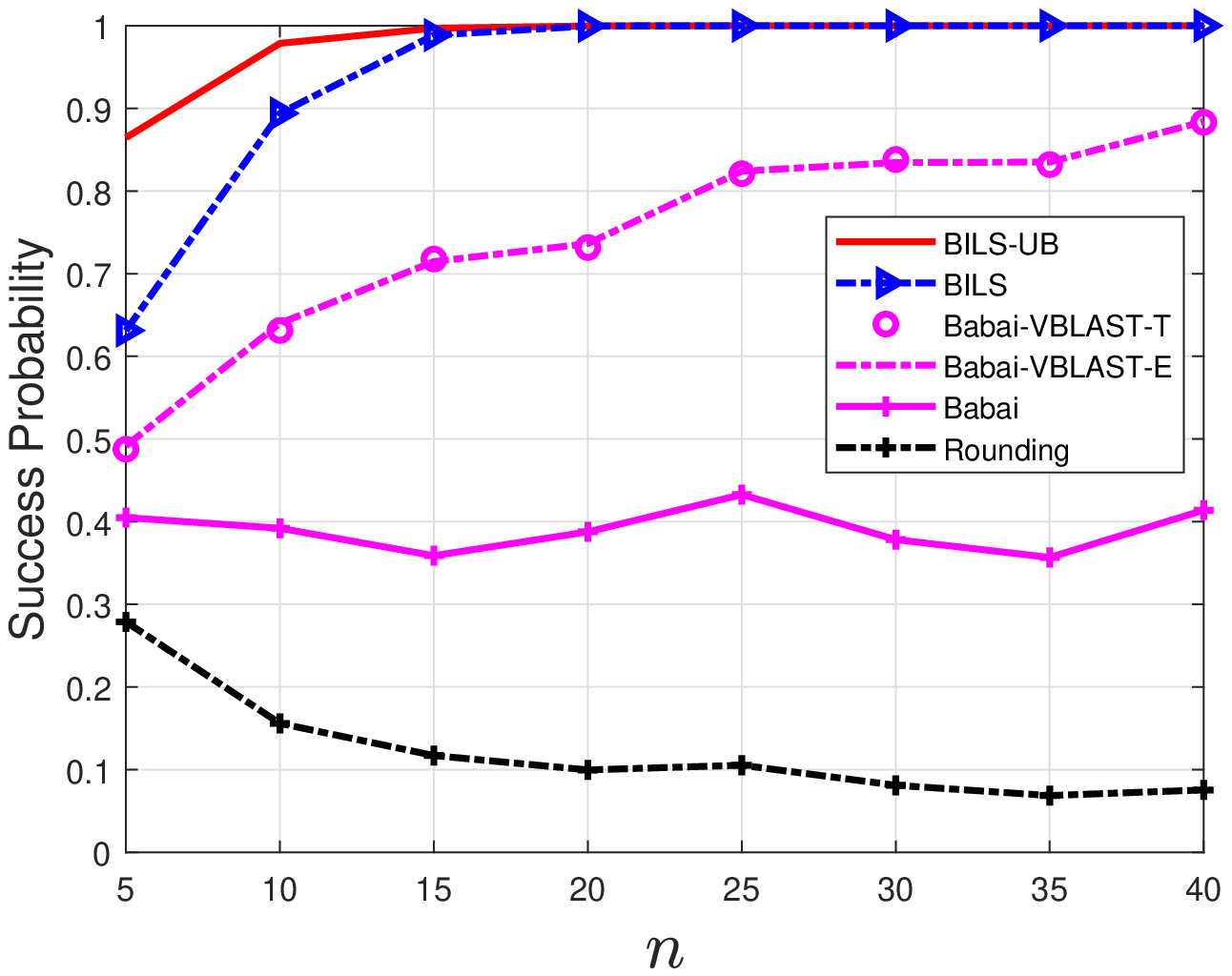}
\caption{Success probabilities versus $n=5:5:40$
for SNR=15 dB and $\mathcal{B}=[0,7]^n$}
\label{fig:n2}
\end{figure}

Figures \ref{fig:sigma1} and \ref{fig:sigma2} display the test results
for SNR$=4:4:32$  dB, $n=20$ with $\mathcal{B}=[0,1]^n$ and $\mathcal{B}=[0,7]^n$,
respectively.
Figures \ref{fig:n1} and \ref{fig:n2} show the test results for $n=5:5:40$, SNR=15 dB with
$\mathcal{B}=[0,1]^n$ and $\mathcal{B}=[0,7]^n$, respectively.

From Figures \ref{fig:sigma1}-\ref{fig:n2}, one can see that the experimental success probability
of $\x^\sBR$ is less than that of $\x^\sBB$, which is less than that of the V-BLAST aided $\x^\sBB$,
and $\x^\sBL$ has the highest success probability.
These observations are consistent with the inequality $P_\sR^\sBR \leq P_\sR^\sBB$ (see \eqref{e:PBRPBB}),
the fact that V-BLAST can improve the success probability of $\x^\sBB$
(more details on this can be found in \cite{WenC17}),
and the inequality $P_\sR^\sBB \leq P_\sR^\sBL$ (see \eqref{e:PBBBL}).
Those figures also show that ``Babai-VBLAST-E" and ``Babai-VBLAST-T" are almost the same,
which means the theoretical $P_\sR^\sBB$ matches very well with the experimental $P_\sR^\sBB$.

Figures \ref{fig:sigma1}-\ref{fig:sigma2} also show that  all the
(experimental) success probabilities of $\x^\sBR$, $\x^\sBB$ (and  the V-BLAST aided $\x^\sBB$),
and $\x^\sBL$ increase as  SNR  increases, and decrease as the box size increases.
Figures \ref{fig:n1}-\ref{fig:n2} show that the success probabilities of all the detectors
decrease when the box size increase for fixed SNR and $n$.
These can easily be explained by Theorems \ref{t:PBRR}, \ref{t:PBBR} and \ref{t:PBLR}, respectively.

In the following, we list observations from Figures \ref{fig:n1}-\ref{fig:n2} for fixed SNR
 and constraint box as $n$ increases,  and give some explanations:
\begin{itemize}
\item The success probability of  the Babai detector $\x^\sBB$ does not change much when $n$ increases.
Since the entries of the tested $\A$ independently and identically follow the standard
Gaussian distribution $\mathcal{N}(0,1)$, by \cite[p99]{Mui82},
the entries of the R-factor $\R$ of the QR factorization (see \eqref{e:qr}) are independent,
and $r_{ii}^2$, $1\leq i\leq n$, follow the Chi-square distribution with
degree $n-i+1$ (so the mean of $r_{ii}^2$ is $n-i+1$), and $r_{ij}$,
$1\leq i\neq j\leq n$, follow the standard Gaussian distribution.
Suppose that $n$ increases to (say) $n'$ and we denote the new R-factor by $\R'$.
Roughly speaking,  the $n'-n$ leading diagonal entries of $\R'$ are large
and  the rest are more or less the same as the diagonal elements of $\R$.
From \eqref{e:PBBR}, we see that after $n$ increases to $n'$,
the product in the formula of $P_\sR^\sBB$ for $\R'$ has $n'-n$ more factors,
corresponding to the $n'-n$ leading diagonal entries of  $\R'$.
Our numerical test indicated that each of these $n'-n$ factors is close to 1.
The rest $n$ factors in the  product are more or less same as those in the product corresponding to $\R$.
This explains why $P_\sR^\sBB$ does not change much when $n$ increases.

\item The success probability of the V-BLAST aided Babai detector  $\x^\sBB$ increases.
Generally speaking,  applying V-BLAST will increase the smallest diagonal  entries of $\R$
and decrease largest ones (note that $\det(\R)$ is unchanged),  i.e., the gap between
the largest one and the smallest one decreases.
This leads to the increase of $P_\sR^\sBB$; see \cite{WenC17} for more details.
For the sake of convenience we denote the new R-factor after applying V-BLAST to $\R$ by $\R_{\sss \mathrm{V}}$
and the new R-factor after applying V-BLAST to $\R'$ (defined in the preceding item for dimension $n'$) by $\R'_{\sss \mathrm{V}}$.
From  the preceding item, the $n$ diagonal entries of $\R'$ are more or less the same as
those of $\R$, but $\R'$ has $n'-n$ extra large entries.
Roughly speaking,  the first $n$ diagonal entries of $\R'_{\sss \mathrm{V}}$
are larger than the $n$ diagonal entries of  $\R_{\sss \mathrm{V}}$.
The rest $n'-n$ diagonal entries of $\R'_{\sss \mathrm{V}}$ are large.
From \eqref{e:PBBR} we see the formula for $P_\sR^\sBB(\R'_{\sss \mathrm{V}})$  involves a product of $n'$ factors,
in which the $n$ factors are larger than the $n$ factors in the product involved in the formula
for $P_\sR^\sBB(\R_{\sss \mathrm{V}})$ and the rest $n'-n$ factors are close to 1 for the tested cases.
Therefore, $P_\sR^\sBB(\R'_{\sss \mathrm{V}})$ is larger than $P_\sR^\sBB(\R_{\sss \mathrm{V}})$.

\item  The upper bound $\mu^\sBL$ on $P_\sR^\sBL$ increases when $n$ increases.
From the above explanation, when $n$ increases to $n'$, denote the new R-factor by
$\R'$, then the $n'-n$ extra terms in \eqref{e:PBLRUB} are very close to 1.
Furthermore, since $\R'$ has more rows than $\R$ and
their nondiagonal entries follow  the same distribution,
$\|\R'\e_j\|^2/r'_{jj}$ for $1\leq j\leq n$ are usually larger than $\|\R\e_i\|^2/r_{ii}$ for $1\leq i\leq n$, so from \eqref{e:PBLRUB},
we can see that $\mu^\sBL$ increases with $n$.
\end{itemize}

To clearly see how good the upper bound given in Theorem \ref{t:PBLRUB} on
the success probability $P_\sR^\sBL$ of $\x^\sBL$,
we display the ratio of ``BILS-UB'' to ``BILS'' versus SNR $=4:4:32$ dB with $n=20$, $\mathcal{B}=[0,1]^n$ and $\mathcal{B}=[0,7]^n$ in Table \ref{tb:ratio1},
and display the ratio of ``BILS-UB'' to ``BILS'' versus $n=5:5:40$ with SNR=15 dB, $\mathcal{B}=[0,1]^n$ and $\mathcal{B}=[0,7]^n$ in Table \ref{tb:ratio2}.
From these two tables, we can see that ``BILS-UB'' is close to ``BILS'' for high SNR or small box,
so the upper bound $\mu^\sBL$ given in Theorem \ref{t:PBLRUB} on the success probability $P_\sR^\sBL$
is sharp for high SNR or small box.

\begin{table*}[htbp!]
\caption{The ratio of the average of the upper bound $\mu^\sBL$ on $P_\sR^\sBL$ given
in Theorem \ref{t:PBLRUB} to  the experimental success probability of $\x^\sBL$ for $n=20$}
\centering
\begin{tabular}{|c||c|c|c|c|c|c|c|c|c|}
\hline
\backslashbox{box}{SNR (dB)} & 4  & 8 & 12 &16 &20 &24&28&32 \\ \hline
$ [0,1]^n$&    1.0000 &    1.0000  &  1.0000 &  1.0000  &  1.0000 &   1.0000 & 1.0000 &   1.0000\\ \hline
$ [0,7]^n$&   55.1833  &  2.8053   & 1.0406 &   1.0000 &    1.0000 &   1.0000 &   1.0000  &  1.0000\\ \hline
\end{tabular}
\label{tb:ratio1}
\end{table*}

\begin{table*}[htbp!]
\caption{The ratio of the average of the upper bound $\mu^\sBL$ on $P_\sR^\sBL$ given
in Theorem \ref{t:PBLRUB} to  the experimental success probability of $\x^\sBL$ for SNR=15 \lowercase{d}B}
\centering
\begin{tabular}{|c||c|c|c|c|c|c|c|c|c|}
\hline
\backslashbox{box}{$n$} & 5  & 10 & 15 &20 &25 &30&35&40 \\ \hline
$ [0,1]^n$&    1.0001  &  1.0000 &   1.0000 &   1.0000 &  1.0000 &   1.0000 &   1.0000  &  1.0000\\ \hline
$ [0,7]^n$&   1.3697  &  1.0947 &   1.0085 &   1.0001  &  1.0000 &   1.0000 &   1.0000  &  1.0000\\ \hline
\end{tabular}
\label{tb:ratio2}
\end{table*}

\section{Conclusion} \label{s:sum}
In this paper, we have investigated the success probabilities of the box-constrained
rounding detectors $\x^\sBR$, the box-constrained Babai detectors $\x^\sBB$
and the BILS detectors $\x^\sBL$ for
detecting an integer parameter vector $\hbx\in \mathcal{B}$ in the linear model \eqref{e:model},
and studied their relationships for two cases:
$\hbx$ is deterministic and $\hbx$ is uniformly distributed over $\mathcal{B}$.
We first developed formulas for the success probabilities  $P_\sD^\sBR$ and $P_\sR^\sBR$
for $\x^\sBR$, the success probability $P_\sD^\sBB$ for $\x^\sBB$,
and the success probabilities  $P_\sD^\sBL$ and $P_\sR^\sBL$ for $\x^\sBL$.
Since it is time consuming to compute $P_\sD^\sBL$ and $P_\sR^\sBL$,
we also developed upper bounds, which can easily be calculated, on them.
Then, we gave two examples to show that
both $P_\sD^\sBR>P_\sD^\sBB$ and $P_\sD^\sBB>P_\sD^\sBL$ are possible,
and rigorously proved that $P_\sR^\sBR\leq P_\sR^\sBB$ always holds.

In MIMO applications, often the entries of $\A$ are assumed
to independently and identically follow the standard Gaussian distribution (see, e.g., \cite{HasV05}).
A closed-form expression of $P_\sR^\sBB$ has been developed in \cite{WenWT17} and \cite{WenWTF18}
for this class of random matrices $\A$.
Although we do not have a formula for $P_\sR^\sBR$ or $P_\sR^\sBL$ for random $\A$,
we can see that $P_\sR^\sBR\leq P_\sR^\sBB \leq P_\sR^\sBL$ still hold
if the entries of $\A$, $\v$ and $\hbx$ are independent random variables,
since it holds for any realization of $\A$ which implies it also holds for random matrix $\A$.

\appendices
\section{Proof of Theorem~\ref{t:PBRR}}
\label{ss:pfPBRR}
\begin{IEEEproof}  Note that \eqref{e:PBRR11} is just the expansion of \eqref{e:PBRR1}.
Thus we need only to prove  \eqref{e:PBRR} and  \eqref{e:PBRR1}.

Since $\hbx$ is uniformly distributed over $\mathcal{B}$, for any $\bar{\x}\in \mathcal{B}$,
$$\Pr(\hbx=\bar{\x})=\frac{1}{\prod_{i=1}^n(u_i-\ell_i+1)}.$$
Thus,
\begin{align}
\Pr(\x^\sBR=\hbx) &= \sum_{\forall \bar{\x}\in\mathcal{B}} \Pr(\x^\sBR = \hbx| \hbx=\bbx) \Pr(\hbx=\bbx)\nonumber\\
&= \sum_{\forall \bar{\x}\in\mathcal{B}} \Pr(\x^\sBR =\bbx) \Pr(\hbx=\bbx) \nonumber\\
&= \sum_{\forall \bar{\x}\in\mathcal{B}} P_\sD^\sBR(\bar{\x}) \Pr(\hbx=\bbx)\nonumber\\
&= \frac{1}{\prod_{i=1}^n(u_i-\ell_i+1)} \sum_{\forall \bar{\x}\in\mathcal{B}} P_\sD^\sBR(\bar{\x}).
\label{e:PBRR}
\end{align}

For any $\bbx\in {\cal B}$, we denote
\[
\bbx = \bmx \tilde{\x} \\ \bar x_n\emx, \ \ \mathcal{\tilde{B}}=\{\tilde{\x} \in \mathbb{Z}^{n-1}:\l_{1:n-1} \leq \tilde{\x} \leq \u_{1:n-1}\},\,
\]
where $\l$ and $\u$ are defined in  \eqref{e:box}.
To simplify notation, we also denote
\[
\gamma_i=u_i-\ell_i+1, \ \  f(\bxi)=\exp\left(-\frac{\|\R\bxi\|^2_2}{2\sigma^2}\right).
\]
Then,  by Theorem \ref{t:PBRD} and \eqref{e:PBRR}, we see  that to show \eqref{e:PBRR1} we only need to show
\begin{align}
&\frac{1}{\prod_{i=1}^n\gamma_i}\sum_{\forall \bar{\x}\in\mathcal{B}}
\int_{{\cal I}_n( \bar{\x})}\cdots \int_{{\cal I}_1( \bar{\x})} f(\bxi)d\xi_1\cdots  d\xi_n\nonumber\\
=&\left(\alpha_n  \int_{-\infty}^{\infty}d\xi_n +\beta_n d\xi_n  \int_{-1/2}^{1/2} \right) \cdot   \cdots   \cdot\nonumber\\
&\times \left( \alpha_1 \int_{-\infty}^{\infty} d\xi_1+\beta_1  \int_{-1/2}^{1/2}d\xi_1 \right)
 f(\bxi).
\label{e:PBRRproof}
\end{align}

By  the fact that $\bar x_n$ can be any integer in $[l_n,u_n]$ and \eqref{e:I}, we have
\begin{align*}
&\frac{1}{\gamma_n}\sum_{\forall \bar{\x}\in\mathcal{B}} \int_{{\cal I( \bar{\x})}_n}\cdots \int_{{\cal I( \bar{\x})}_1}
f(\bxi)d\xi_1\cdots d\xi_n \nonumber\\
=&\frac{1}{\gamma_n}\sum_{\forall \tilde{\x}\in\mathcal{\tilde{B}}}\left[\int_{-\infty}^{1/2} \int_{{\cal I( \tilde{\x})}_{n-1}}\cdots \int_{{\cal I( \tilde{\x})}_1}
f(\bxi)d\bxi \right.\nonumber\\
&+\int^{\infty}_{-1/2} \int_{{\cal I( \tilde{\x})}_{n-1}}\cdots \int_{{\cal I( \tilde{\x})}_1}f(\bxi)d\bxi \nonumber\\
&+\left.(u_n-\ell_n-1)\int^{1/2}_{-1/2} \int_{{\cal I( \tilde{\x})}_{n-1}}\cdots \int_{{\cal I( \tilde{\x})}_1}f(\bxi)d\bxi \right]\nonumber\\
=&\frac{1}{\gamma_n}\sum_{\forall \tilde{\x}\in\mathcal{\tilde{B}}}
\left[\int_{-\infty}^{1/2}+\int^{\infty}_{-1/2} +(u_n-\ell_n-1)\int^{1/2}_{-1/2}\right]\nonumber\\
&\int_{{\cal I( \tilde{\x})}_{n-1}}\cdots \int_{{\cal I( \tilde{\x})}_1}f(\bxi)d\bxi \nonumber\\
=&\sum_{\forall \tilde{\x}\in\mathcal{\tilde{B}}}\left(\alpha_n  \int_{-\infty}^{\infty} +\beta_n  \int_{-1/2}^{1/2} \right)
\int_{{\cal I( \tilde{\x})}_{n-1}}\cdots \int_{{\cal I( \tilde{\x})}_1}f(\bxi)d\bxi.
\end{align*}
By repeating the above procedure $n-1$ times, we obtain \eqref{e:PBRRproof}.

\end{IEEEproof}

\section{Proof of Theorem \ref{t:PBLD}}
\label{ss:pfPBLD}
\begin{IEEEproof}
Define sets ${\cal S}_1, {\cal S}_2$ and ${\cal S}_3$ as
\begin{align}
\label{e:S1}
{\cal S}_1  =\{& \tilde{\v} \,|\,  \tilde{\v}\sim {\cal N}(\0,\sigma^2\I), \
 \tby = \R\hbx + \tilde{\v}, \  \x^\sBL=\hbx\},\\
\label{e:S2}
{\cal S}_2  =\{&  \tilde{\v} \,|\, \tilde{\v}\sim {\cal N}(\0,\sigma^2\I), \
 \tby = \R\hbx + \tilde{\v}, \nonumber\\
 & \|\tilde{\v}\|_2^2 \leq \| \tby-\R\x\|_2^2 \mbox{ for } \forall \,\x\in\mathcal{B}\}, \\
{\cal S}_3 =\{ &\tilde{\v} \,|\, \tilde{\v}\sim {\cal N}(\0,\sigma^2\I), \ \tby = \R\hbx + \tilde{\v},\nonumber\\
& \| \tilde{\v}\|_2^2 = \| \tby-\R\x\|_2^2 \mbox{ for some } \x\neq \hbx,  \x\in\mathcal{B} \}. \label{e:S3}
\end{align}
Then, we can easily see that
\beq
\label{e:BLvconstraint}
{\cal S}_2\backslash {\cal S}_3\subseteq {\cal S}_1\subseteq {\cal S}_2.
\eeq
Thus,
\beq
\label{e:PS1S2}
\Pr(\tilde{\v}\in {\cal S}_2)-\Pr(\tilde{\v}\in {\cal S}_3)\leq P_\sD^\sBL\leq \Pr(\tilde{\v}\in {\cal S}_2).
\eeq

In the following we show
\beq
\label{e:PS2}
\Pr(\tilde{\v}\in {\cal S}_3)=0.
\eeq
Let $\tilde{\v}\in {\cal S}_3$, then there exists at least one $\bar{\x}\in \mathcal{B}$
such that $\bar{\x}\neq \hbx$ and $\| \tilde{\v}\|_2^2 = \| \tby-\R\bar{\x}\|_2^2$.
Note that by \eqref{e:modelqr}, we have $\tilde{\v} =\tby-\R\hat{\x}$.
Then, with \eqref{e:modelqr}, we have
\begin{align*}
\| \tilde{\v}\|_2^2&=\| \tby-\R\bar{\x}\|_2^2
=\| (\tby-\R\hbx)-\R(\bar{\x}-\hbx)\|_2^2  \nonumber \\
&=\| \tilde{\v} \|_2^2+\|\R(\bar{\x}-\hbx)\|_2^2-2(\bar{\x}-\hbx)^T\R^T\tilde{\v}.
\end{align*}
Thus
\[
2(\bar{\x}-\hbx)^T\R^T\tilde{\v} \!= \!  \|\R(\bar{\x}-\hbx)\|_2^2
\]
which indicates that  $\tilde{\v}$ lies on an $(n-1)-$dimensional plane.
Since $\tilde{\v}$ is an $n-$dimensional Gaussian random variable, \eqref{e:PS2} holds.

By \eqref{e:PS1S2}-\eqref{e:PS2}, we can see that
\beq
\label{e:PV}
P_\sD^\sBL=\Pr(\tilde{\v}\in {\cal S}_2).
\eeq
For $\tilde{\v} \in {\cal S}_2$ and $\x\in\mathcal{B}$,
\begin{align*}
\| \tby-\R\x\|_2^2
&=\| (\tby-\R\hbx)-\R(\x-\hbx)\|_2^2\\
&=\| \tilde{\v} \|_2^2+\|\R(\x-\hbx)\|_2^2-2(\x-\hbx)^T\R^T\tilde{\v},
\end{align*}
which implies that $\| \tilde{\v}\|_2^2 \leq \| \tby-\R\x\|_2^2$ if and only if
\[
2(\x-\hbx)^T\R^T\tilde{\v} \leq \|\R(\x-\hbx)\|_2^2.
\]
Then, from \eqref{e:PV} and the fact that  $\tilde{\v}\sim {\cal N}(\0,\sigma^2\I)$,
we can conclude that \eqref{e:PBLD} holds.
\end{IEEEproof}

\section{Proof of Theorem~\ref{t:PBLDUB}}
\label{ss:pfPBLDUB}

\begin{IEEEproof}
Let $\x$ be any element in the set ${\cal B}$ that is not equal to $\hbx$.
Denote
$$
{\cal S}_\x = \{\bxi\mid  2(\x-\hbx)^T\R^T\bxi \!\leq \!  \|\R(\x-\hbx)\|_2^2\}.
$$
Then by \eqref{e:S}, ${\cal S^\sBL \in {\cal S}}_\x$.
Thus, by \eqref{e:PBLD},
\beq \label{e:pdbl_ine}
P_\sD^\sBL \leq
\frac{1}{(\sqrt{2\pi } \sigma)^{n}}\int_{{\cal S}_\x}\exp\left(-\frac{1}{2\sigma^2}\|\bxi\|_2^2\right)d\bxi.
\eeq

Let $\W \in \Rnbn$ be an orthogonal matrix (e.g., the Householder matrix) such that
\beq \label{eq:orthtrans}
\W \R(\x-\hbx) = \|\R(\x-\hbx)\|_2 \e_1.
\eeq
(Recall $\e_1$ is the first column of the $n\times n$ identity matrix.)
Since $\x\neq \hbx$, $\|\R(\x-\hbx)\|_2\neq 0$ (note that $\R$ is nonsingular).
Set $\boldeta:=\W \bxi$. Then  ${\cal S}_\x$ is transformed to
$$
\bar{\cal S}_\x =   \{\boldeta  \mid  \eta_1  \!\leq \!   \|\R(\x-\hbx)\|_2/2\}.
$$
Then from \eqref{e:pdbl_ine}, we obtain
\begin{align*}
P_\sD^\sBL& \leq
 \frac{1}{(\sqrt{2\pi } \sigma)^{n}}\int_{\bar{\cal S}_\x} \exp\left(-\frac{1}{2\sigma^2}\|\boldeta\|_2^2\right)d\boldeta \\
&=\frac{1}{\sqrt{2\pi } \sigma}
\int_{-\infty}^{ \frac{1}{2} \|\R(\x-\hbx)\|_2}\exp\big(-\frac{ \eta_1^2}{2\sigma^2})d\eta_1\\
&\times\frac{1}{(\sqrt{2\pi } \sigma)^{n-1}}\int_{\mathbb{R}^{n-1}}\exp\big(-\frac{1}{2\sigma^2}\|\boldeta_{2:n}\|_2^2)d\boldeta_{2:n}\\
&=\frac{1}{2}\big[1+\phi_\sigma( \|\R(\x-\hbx)\|_2)\big],
\end{align*}
where the last equality follows from \eqref{e:varphi}, hence \eqref{e:PBLDUB1} holds.

Let $\min\limits_{1\leq i\leq n}\|\R_{1:i,i}\|_2=\|\R_{1:j,j}\|_2$.
Note that  $\hbx+\e_j\in {\cal B}$ or $\hbx-\e_j\in {\cal B}$.
Suppose that $\hbx+\e_j\in {\cal B}$, then \eqref{e:PBLDUB} follows from \eqref{e:PBLDUB1}
by setting $\x=\hbx+\e_j$.
Otherwise, \eqref{e:PBLDUB} follows from \eqref{e:PBLDUB1} by setting $\x=\hbx-\e_j$.
 \end{IEEEproof}

\section{Proof of Lemma~\ref{l:integralineq1}}
\label{ss:pfintegralineq1}

\begin{IEEEproof}
An inequality which is equivalent to \eqref{e:integralineq1} for $a_i=1/2$ is given in \cite{Teu98a}.
As a reviewer pointed out, when all $a_i$ are finite,
we could prove  \eqref{e:integralineq1} by applying that inequality via a change of variables.
But for the reader's convenience, we  give a proof without referring to \cite{Teu98a}.

We prove the lemma by changing variables in the integral.
Let
\[
\T =\bmx
1& - \frac{1}{u_{11}} \U_{1,2:n} \\
\0& \I_{n-1}
\emx,
\]
then
$$
\U\T = \bmx u_{11} & 0 \\ \0 & \U_{2:n,2:n} \emx.
$$
Define $\bxi=\T \boldsymbol{\eta}$, i.e.,
$$
\xi_1 = \eta_1 - \frac{1}{u_{11}}\U_{1,2:n}\boldsymbol{\eta}_{2:n}, \ \ \bxi_{2:n}=\boldsymbol{\eta}_{2:n}.
$$

First we consider the case that $a_1< \infty$.
Since $\xi_1 \in [-a_1,a_1]$,
\[
\eta_1 \in s_1:= \Big[- a_1 + \frac{1}{u_{11}}\U_{1,2:n}\boldsymbol{\eta}_{2:n},\   a_1+  \frac{1}{u_{11}}\U_{1,2:n}\boldsymbol{\eta}_{2:n}\Big].
\]
Then we have
\begin{align*}
&  \int_{-a_n}^{a_n}\cdots  \int_{-a_1}^{a_1}
\exp\left(- \|\U\bxi\|^2 \right) d\xi_{1} \cdots d \xi_n \nonumber\\
=&  \int_{-a_n}^{a_n}\cdots  \int_{s_1}
\exp(- u_{11}^2\eta^2_1- \|\U_{2:n,2:n}\boldsymbol{\eta}_{2:n}\|_2^2) d\eta_{1} \cdots d \eta_n \nonumber\\
=&  \int_{s_1}\exp(- u_{11}^2\eta^2_1) d\eta_1\\
&\times \int_{-a_n}^{a_n}\cdots  \int_{-a_2}^{a_2}
\exp(- \|\U_{2:n,2:n}\boldsymbol{\eta}_{2:n}\|_2^2) d \eta_{2} \cdots d \eta_n.
\end{align*}

According to \cite[eq.\ (68)]{WenC17}, we have
\beq
 \label{e:Gaussianine}
\int_{s_1}\exp\left(- u_{11}^2\eta^2_1 \right)d\eta_1
\leq \int_{-a_1}^{a_1}\exp\left(- u_{11}^2t^2 \right)dt,
\eeq
which actually can easily be observed from the graph of the density function of the normally distributed random variable with 0 mean.
Therefore,
\begin{align}
\label{e:integralineq2}
&  \int_{-a_n}^{a_n}\cdots  \int_{-a_1}^{a_1}
\exp\left(- \|\U\bxi\|^2 \right) d\eta_{1} \cdots d \eta_n \nonumber\\
\leq &  \int_{-a_1}^{a_1}\exp\left(- u_{11}^2t^2 \right)dt \nonumber \\
& \times \int_{-a_n}^{a_n}\cdots  \int_{-a_2}^{a_2}
 \exp(- \|\U_{2:n,2:n}\boldsymbol{\eta}_{2:n}\|_2^2) d\eta_{2} \cdots d \eta_n.
\end{align}
Thus, by \eqref{e:integralineq2}, one can easily show that
\begin{align*}
&  \int_{-a_n}^{a_n}\cdots  \int_{-a_1}^{a_1}
\exp\left(- \|\U\bxi\|^2 \right) d\eta_{1} \cdots d \eta_n \\
\leq &  \int_{-a_1}^{a_1}\exp\left(- u_{11}^2t^2 \right)dt
\int_{-a_2}^{a_2}\exp\left(- u_{11}^2t^2 \right)dt\\
& \times \int_{-a_n}^{a_n}\cdots  \int_{-a_3}^{a_3}
 \exp(- \|\U_{3:n,3:n}\boldsymbol{\eta}_{3:n}\|_2^2) d\eta_{3} \cdots d \eta_n\\
\leq&\ldots\leq\prod_{i=1}^n \int_{-a_i}^{a_i}\exp\left(- u_{ii}^2t^2 \right)dt.
\end{align*}
Hence, \eqref{e:integralineq1} holds for finite $a_i,1\leq i\leq n$.

In the following, we show that \eqref{e:integralineq1} holds if some or all $a_i$ are infinity.
To show this, by the above analysis, it suffices to show that
\eqref{e:integralineq2} still holds if $a_1=\infty$.
Note  that
\begin{align*}
& \int_{-a_n}^{a_n}\cdots \int_{-\infty}^\infty
\exp\left(-\|\U\bxi\|_2^2 \right)d\xi_{1}\cdots d \eta_n \\
=\ & \lim_{\alpha \rightarrow \infty,\beta \rightarrow \infty}  \int_{-a_n}^{a_n}\cdots \int_{-\alpha}^{\beta}
\exp\left(- \|\U\bxi\|_2^2 \right)d\xi_{1}\cdots d \eta_n \\
=\ & \lim_{\alpha \rightarrow \infty }  \int_{-a_n}^{a_n}\cdots \int_{-\alpha}^{\alpha}
\exp\left(- \|\U\bxi\|_2^2 \right)d\xi_{1} \cdots d\xi_n,
\end{align*}
where  the second equality is due to the fact that the improper integral is convergent.
Then taking  limit on both sides of \eqref{e:integralineq2} as $\alpha \rightarrow \infty$
leads to the desired inequality.
\end{IEEEproof}

\section{Proof of Theorem~\ref{t:PBLRUB}}
\label{ss:pfPBLRUB}

\begin{IEEEproof}
To prove \eqref{e:PBLRUB}, we first prove the following inequality
\begin{align}
\label{e:pdbu_ine}
P_\sD^\sBL(\hbx)  &\leq
\frac{\det(\R^{-T})}{(\sqrt{2\pi } \sigma)^{n}}\times\nonumber\\
& \int_{{\cal I}_1(\hbx)}\cdots \int_{{\cal I}_n(\hbx)}
\exp\left(-\frac{1}{2\sigma^2}\|\R^{-T}\boldeta\|_2^2\right)d\eta_n \cdots d\eta_1,
\end{align}
where
\beq
\label{e:Ihbx}
{\cal \hat{I}}(\hbx):={\cal \hat{I}}_1(\hbx)\times {\cal \hat{I}}_2(\hbx)  \times \cdots \times {\cal \hat{I}}_n(\hbx)
\eeq
with
\beq
\label{e:Ihbxi}
{\cal \hat{I}}_i(\hbx) :=
\begin{cases}
(-\infty,  \|\R \e_i \|_2^2/2], & \hx_i = \ell_i \\
(-  \|\R\e_i \|_2^2/2, \|\R\e_i \|_2^2/2 ],  & \ell_i < \hx_i < u_i \\
(-  \|\R\e_i \|_2^2/2, \infty), & \hx_i = u_i
\end{cases}
\eeq
for $1\leq i\leq n$.

By Theorem \ref{t:PBLD} and setting $\boldeta = \R^T\bxi$, we have
\beq
 \label{eq:pdbl}
P_\sD^\sBL(\hbx)
=\frac{\det(\R^{-T})}{(\sqrt{2\pi } \sigma)^{n}}\int_{\bar{\cal S}(\hbx)}\exp\big(-\frac{1}{2\sigma^2}\|\R^{-T}\boldeta\|_2^2)d\boldeta,
\eeq
where
\beq
\label{eq:bS}
\bar{\cal S}(\hbx) \!=\!\{\boldeta \mid  (\x-\hbx)^T \boldeta \!\leq \!  \|\R(\x-\hbx)\|_2^2/2 \  \mbox{ for }\forall \,\x\in\mathcal{B}\}.
\eeq
We take some special $\x \in {\cal B}$ so that we will get a set, which is included in
$\bar{\cal S}(\hbx) $, but is more structured so that we can derive an upper bound on $P_\sD^\sBL(\hbx)$,
which can be easily computed. For $1\leq i\leq n$, define
$$
\x^{(i)} =
\begin{cases}
\hbx + \e_i  & \hx_i = \ell_i \\
\hbx \pm \e_i & \ell_i < \hx_i < u_i \\
\hbx -\e_i  & \hx_i = u_i
\end{cases}, \ \
$$
then $\x^{(i)} \in {\cal B}$.
For $1\leq i\leq n$, take $\x=\x^{(i)}$, then the inequality in \eqref{eq:bS} just becomes
$\eta_i \in {\cal \hat{I}}_i(\hbx)$, where ${\cal \hat{I}}_i(\hbx)$ is defined in \eqref{e:Ihbxi}.
Then, by \eqref{e:Ihbx} and \eqref{eq:bS}, $\bar{\cal S}(\hbx) \subseteq {\cal \hat{I}}(\hbx)$.
Therefore, \eqref{e:pdbu_ine} holds.

In the following, we use Theorem \ref{t:PBLR} and \eqref{e:pdbu_ine} to prove \eqref{e:PBLRUB}.
To simplify notation, denote
\[
\gamma_i=u_i-\ell_i+1, \ \  h(\boldeta)=\exp\left(-\frac{\|\R^{-T}\boldeta\|^2_2}{2\sigma^2}\right).
\]
Then, by Theorem \ref{t:PBLR} and \eqref{e:pdbu_ine}, we have
\begin{align*}
P_\sR^\sBL\leq &\frac{1}{\prod_{i=1}^n\gamma_i}\frac{\det(\R^{-T})}{(\sqrt{2\pi } \sigma)^{n}}\\
&\sum_{\forall \bar{\x}\in\mathcal{B}}
\int_{{\cal \hat{I}}_n( \bar{\x})}\cdots \int_{{\cal \hat{I}}_1( \bar{\x})}
h(\boldeta)d\eta_1\cdots  d\eta_n\nonumber\\
=&\frac{\det(\R^{-T})}{(\sqrt{2\pi } \sigma)^{n}}
\left(\alpha_n  \int_{-\infty}^{\infty}d\eta_n +\beta_n d\eta_n
\int_{-\|\R\e_n \|_2^2/2}^{\|\R\e_n \|_2^2/2} \right) \cdot   \cdots   \cdot\\
&\times\left( \alpha_1 \int_{-\infty}^{\infty} d\eta_1+\beta_1  \int_{-\|\R\e_1 \|_2^2/2}^{\|\R\e_1 \|_2^2/2}d\eta_1 \right) h(\boldeta)\\
\leq &\frac{\det(\R^{-T})}{(\sqrt{2\pi } \sigma)^{n}}
\prod_{i=1}^n \biggl(\alpha_i\int_{-\infty}^\infty \exp\biggl(-\frac{t^2}{2\sigma^2r_{ii}^2}\biggr)dt\\
&+ \beta_i \int_{-\|\R\e_i\|_2^2/2}^{\|\R\e_i\|_2^2/2}
      \exp\biggl(-\frac{t^2}{2\sigma^2r_{ii}^2}\biggr)dt \biggr) \\
=\ &   \prod_{i=1}^n \biggl( \frac{1}{u_i-\ell_i+1}
+ \frac{u_i-\ell_i}{u_i-\ell_i+1} \phi_\sigma\biggl(\frac{\|\R\e_i\|_2^2}{r_{ii}}\biggr)\biggr),
\end{align*}
where $\alpha_i$ and $\beta_i$ are defined in \eqref{e:alphabeta},
the first equality is obtained by using similar method for showing \eqref{e:PBRRproof};
the second inequality follows from using Lemma \ref{l:integralineq1} with $\U=\R^{-T}/\sqrt{2\sigma}$
to each item of the expansion of the right-hand side of the first equality;
and the second equality is from \eqref{e:varphi} and the integral transformation.
\end{IEEEproof}

\section{Proof of Theorem~\ref{t:PBRPBB}}
\label{ss:PBRPBB}
\begin{IEEEproof}
Applying Lemma \ref{l:integralineq1} by taking $\U=\R/(\sqrt{2}\sigma)$
to each term in the sum in   \eqref{e:PBRR11},  which is the expanded version
of \eqref{e:PBRR1},  and then combining all the terms into the same form as \eqref{e:PBRR1}, we obtain
\begin{align*}
P_\sR^\sBR
\leq &
\frac{\det(\R)}{(\sqrt{2\pi} \sigma)^{n}}  \prod_{i=1}^n
\left( \frac{1}{u_i-\ell_i+1} \int_{-\infty}^{\infty} \exp\left(-\frac{r_{ii}^2}{2\sigma^2}t^2\right) dt  \right.   \\
& \ \ + \left. \frac{u_i-\ell_i}{u_i-\ell_i+1} \int_{-1/2}^{1/2}  \exp\left(-\frac{r_{ii}^2}{2\sigma^2}t^2\right) dt \right).
\end{align*}
Then using the facts that
$$
\frac{r_{ii}}{\sqrt{2\pi} \sigma}  \int_{-\infty}^{\infty} \exp\left(-\frac{r_{ii}^2}{2\sigma^2}t^2\right) dt =1
$$
and (see \eqref{e:varphi})
$$
\frac{r_{ii}}{\sqrt{2\pi} \sigma}  \int_{-1/2}^{1/2} \exp\left(-\frac{r_{ii}^2}{2\sigma^2}t^2\right) dt = \phi_\sigma(r_{ii})
$$
we have
$$
P_\sR^\sBR
\leq \prod_{i=1}^n \Big[\frac{1}{u_i-\ell_i+1} + \frac{u_i-\ell_i}{u_i-\ell_i+1} \phi_\sigma(r_{ii}) \Big],
$$
where the right-hand side is just $P_\sR^\sBB$ by Theorem \ref{t:PBBR}.
\end{IEEEproof}

\bibliographystyle{IEEEtran}
\bibliography{ref}
\end{document}